\documentclass[aps,pre,reprint,showpacs,showkeys]{revtex4-1}
\usepackage{epsfig}
\usepackage{amsmath}
\usepackage{multirow,bigdelim}
\usepackage{booktabs} %for making horizontal lines for only a few columns

\usepackage{color}

\usepackage{subfig}
\usepackage{caption}
\DeclareCaptionFormat{center}%{{\centerline{#1#2}\\ }#3}

\newlength\replength
\newcommand\repfrac{.33}

\newcommand\rulewidth{.6pt}
\newcommand\tdashfill[1][\repfrac]{\cleaders\hbox to \replength{%
  \smash{\rule[\arraystretch\ht\strutbox]{\repfrac\replength}{\rulewidth}}}\hfill}

\newcommand\tdotfill[1][\repfrac]{\cleaders\hbox to \replength{%
  \smash{\raisebox{\arraystretch\dimexpr\ht\strutbox-.1ex\relax}{.}}}\hfill}

\begin{document}
\title{Characterizing the complexity of time series network graphs: A simplicial approach}

\author{N. Nirmal Thyagu}
\affiliation{Division of Physics, \\
School of Advanced Sciences, Vellore Institute of Technology, \\
Chennai Campus, Chennai - 600127, India}

\author{Nithyanand Rao, Malayaja Chutani  and Neelima Gupte}
\affiliation{ Department of Physics,\\
      Indian Institute of Technology Madras,\\
      Chennai - 600036, India\\}

\date{\today}

\begin{abstract}
We analyze the time series obtained from different dynamical regimes of the logistic map by constructing their equivalent time series (TS) networks, using the visibility algorithm. 
The regimes analyzed include both periodic and chaotic regimes, as well as intermittent regimes and the Feigenbaum attractor at the edge of chaos. We use the methods of algebraic topology to define the simplicial characterizers, which can analyse the simplicial structure of the networks at both the global and local levels.
The simplicial characterisers bring out the hierarchical levels of complexity at various topological levels. These hierarchical levels of complexity find the skeleton of the local dynamics embedded in the network which influence the global dynamical properties of the system, and also permit the identification of dominant motifs.
We also analyze the same networks using conventional network characterizers such as average path lengths and clustering coefficients. We see that the simplicial characterizers  are capable of distinguishing between different dynamical regimes and  can pick up subtle differences in dynamical behavior, whereas the usual characterizers provide a coarser characterization. However, the two taken in conjunction, can provide information about the dynamical behavior of the time series, as well as the correlations in the evolving system. Our methods can therefore provide powerful tools for the analysis of dynamical systems. 

\end{abstract}


\maketitle

\section{Introduction}

The analysis of the time series of the variables of evolving dynamical systems is  an important tool for analyzing the dynamical behavior of nonlinear dynamical systems, as well as for making predictions for their behavior. A variety of well developed methods and tools are used to carry out this kind of analysis, which also define a set of  precise metrics such as the Fourier transforms, power spectra, generalized dimensions and entropies, multifractal spectra,  and  Lyapunov exponents \cite{Kantz}.

In recent years, traditional techniques of the analysis of time series have been supplemented by new approaches  which draw on areas like nonlinear time series analysis \cite{TSA_refs}, data mining, and complex networks \cite{GaoEPL2016}. These approaches complement the traditional techniques and provide important additional insights into the behavior of evolving dynamical systems.

An important recent method of analyzing time series, consists of mapping these time series to networks, which are then called time series (TS) networks. A number of methods are available for carrying out this mapping. These include, the use of visibility graphs \cite{vis_algo}, the quantile mapping \cite{Campanharo}, recurrence time networks \cite{Marwan}, etc. These network representations are then analyzed using a variety of well known network metrics such as clustering coefficients, degree distributions  and path lengths \cite{Strogatz98, AlbertBarabasi2002}.

In this paper, we analyze the time series networks (TS networks) obtained from the logistic map at different parameter values, using methods of algebraic topology \cite{Atkin72, Duckstein, Kramer}, such as persistent homology \cite{Carlsson2009, CompTopol}, and recently constructed measures which analyze the simplicial structure of graphs \cite{Andjelkovic1}. The TS networks are constructed from the time series by using the visibility algorithm \cite{vis_algo}, which has certain advantages over other methods. The graphs so constructed are then analyzed using the simplicial characterizers which reveal the hierarchical levels of complexity hidden in the TS network, which arise due to the correlations of the original time series. 

We show that the methods are able to identify the crucial differences between the time series corresponding to distinct dynamical behaviors. We analyze the networks using the usual network characterizers, and demonstrate that the simplicial characterizers, which include both global and local quantities, provide a more sensitive and accurate diagnosis of the dynamical characteristics of the underlying time series.

While the conventional network charactizers give us the global structural information of the network, the local dynamical information is embedded in the simplical complexes and their interconnections at various topological levels. In essence, this formalism uncovers the hidden skeleton of the dynamics of the systems.

We have used multiple metrics that  characterize not only the global properties related to the network structure, but also identifies the hidden skeleton in the form of	hierarchical levels of complexity that expose the underlying dynamics and the short term correlations of the system.
We use the logistic map as a test bed, as its dynamics is well understood, and has been characterized using a variety of conventional quantities such as Lyapunov exponents, entropies etc. We compare the information extracted from  our topological characterizers with that obtained from these quantities, and identify elements that can be generalized to other cases.

We note that the use of characterizers from algebraic topology is slowly finding acceptance in the analysis of complex systems. These include the use of persistent homologies for the topological characterization and early detection of bifurcations \cite{MittalGupta2017, MaleticRajkovic2016}, as well as the analysis of high dimensional data \cite{GiustiGhristBassett2016}. Our methods can also contribute to the effective analysis of these cases.

%--------------------------------------------------------------------------------------------------------------------------------------------
% BIFURCATION DIAGRAM WITH MARKERS
%--------------------------------------------------------------------------------------------------------------------------------------------
\begin{figure}[!t]
	\includegraphics[width=0.50\textwidth]{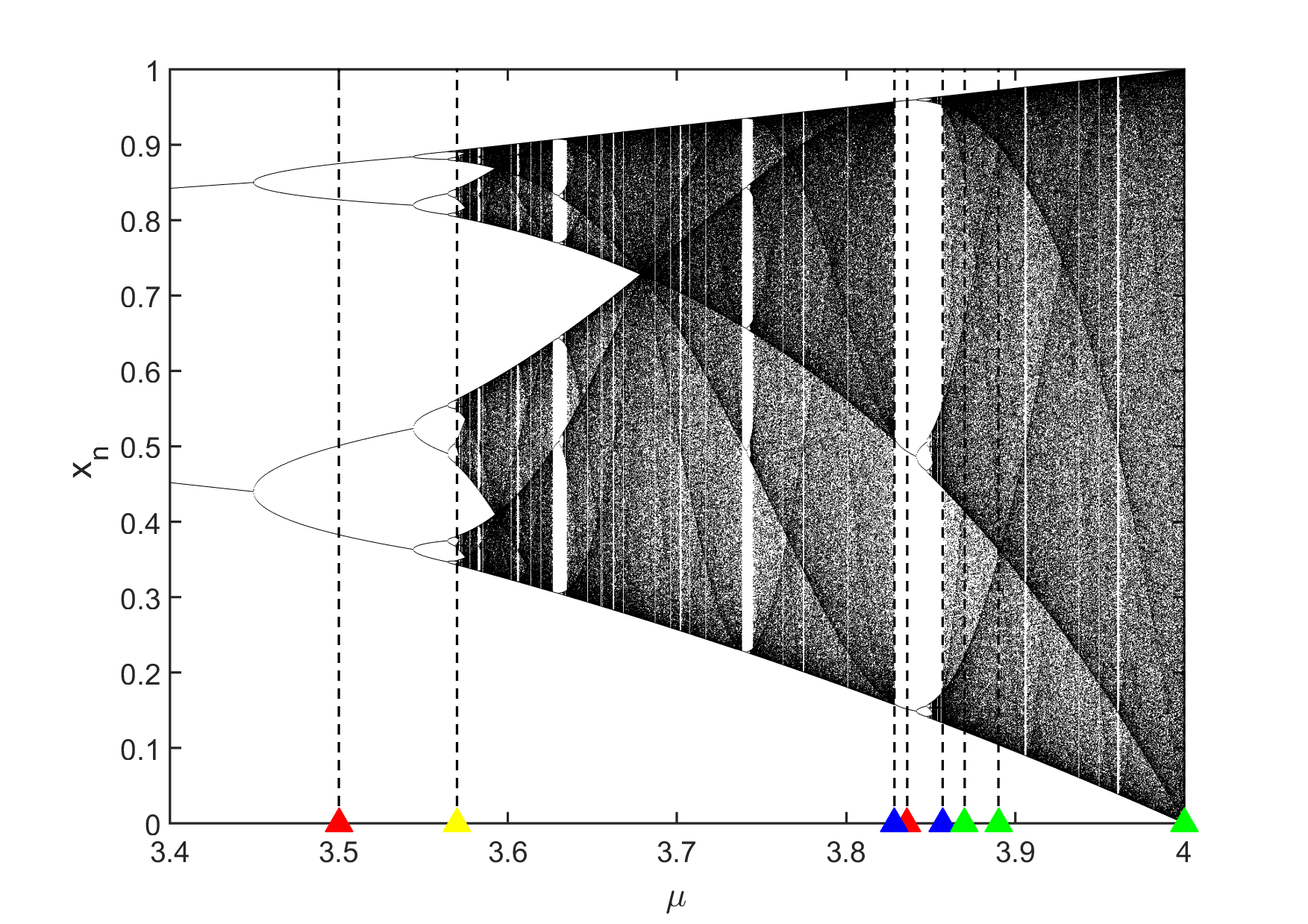}
	\captionsetup{justification=raggedright, singlelinecheck=false}
	\caption{ (Color online) Time-series networks are constructed using the visibility algorithm
for the time series data of the logistic map at different parameter values. 
The bifurcation diagram of the logistic map showing eight parameter
values - periodic: $\mu = 3.5$ and $\mu = 3.836$ (red triangles), 
intermittent: $\mu = 3.8284$ and $\mu = 3.857$ (blue triangles), 
Feigenbaum point: $\mu = 3.56995$ (yellow triangle), 
chaotic: $\mu = 3.87$, $\mu = 3.89$, and $\mu = 3.857$ (green triangles).
\label{logisticMap}}
\end{figure}
%--------------------------------------------------------------------------------------------------------------------------------------------

This paper is organized as follows. We use the time series data from the logistic map at parameter values that show distinct kinds of dynamical behaviors - periodic, chaotic, intermittent, etc. (See Fig. \ref{logisticMap}). To map these time series data sets into their corresponding network representations we employ the visibility algorithm \cite{vis_algo}, as explained in Section \ref{Vgraph}. We define the measures used to analyze the simplicial structures of the resulting graphs, and their connection with  the  topological structure and topological connectivity in Section \ref{Nets}. The results of the analysis are also tabulated in this section. The usual network characterizers of the network are constructed and tabulated in Section \ref{ConventionalCharacs}. We compare the information resulting from these two types of characterizers in, Section \ref{results} and summarize and conclude in Section \ref{conclusion}.

%--------------------------------------------------------------------------------------------------------------------------------------------
%--------------------------------------------------------------------------------------------------------------------------------------------
\section{The Visibility Graph \label{Vgraph}}

In this paper, we use the visibility algorithm developed in Ref.\,\cite{vis_algo} to transform the time series obtained from the logistic map evolving in different dynamical regimes, into a set of networks. Given a time series, visibility graphs are constructed by considering time data points as nodes, and a link is established between any two nodes if and only if there is no obstruction in the line of sight of these two nodes.

Recent developments show that going from the time series representation to the network representation yields additional information of the underlying dynamics \cite{Luque, Campanharo}. While  a number of methods have been developed over recent years to convert a time series into a network \cite{Zhang2006, Yang2008, Campanharo}, we use the visibility algorithm here due to the simplicity of the visibility approach, as well as the advantages it has over other methods viz. the visibility approach does not require any embedding step for phase space reconstruction \cite{SmallSuperfamilyMotifs}, as well as being computationally efficient and analytically tractable \cite{LacasaNonlin, Gutin2011}. Additionally, a TS network resulting from the visibility algorithm conserves the structure of the time series, viz., a periodic time series gives rise to a regular graph, a random time series gives rise to a random graph, and a fractal time series gives rise to a scale-free graph \cite{vis_algo}. We note that the visibility algorithm has been used in diverse contexts ranging from finance to geophysics \cite{Elsner, Yang}.

The visibility algorithm introduced in \cite{vis_algo} is implemented as follows. Let the pair of points $(y_i, t_i)$ denote the data in the time series for $i=1,2,\ldots N$. For any two pairs $(y_i, t_i)$ and $(y_j, t_j)$ to be visible to each other (by line of sight visibility), all other intermediate data pairs $(y_r, t_r)$ should satisfy the condition:

\begin{equation}
	y_j > y_r + \frac{y_j - y_i}{t_j - t_i} (t_j - t_r) \\
\label{visibility}
\end{equation}

\begin{figure}[t]
\begin{center}
\setlength{\unitlength}{0.012500in}%
\includegraphics[width=\columnwidth]{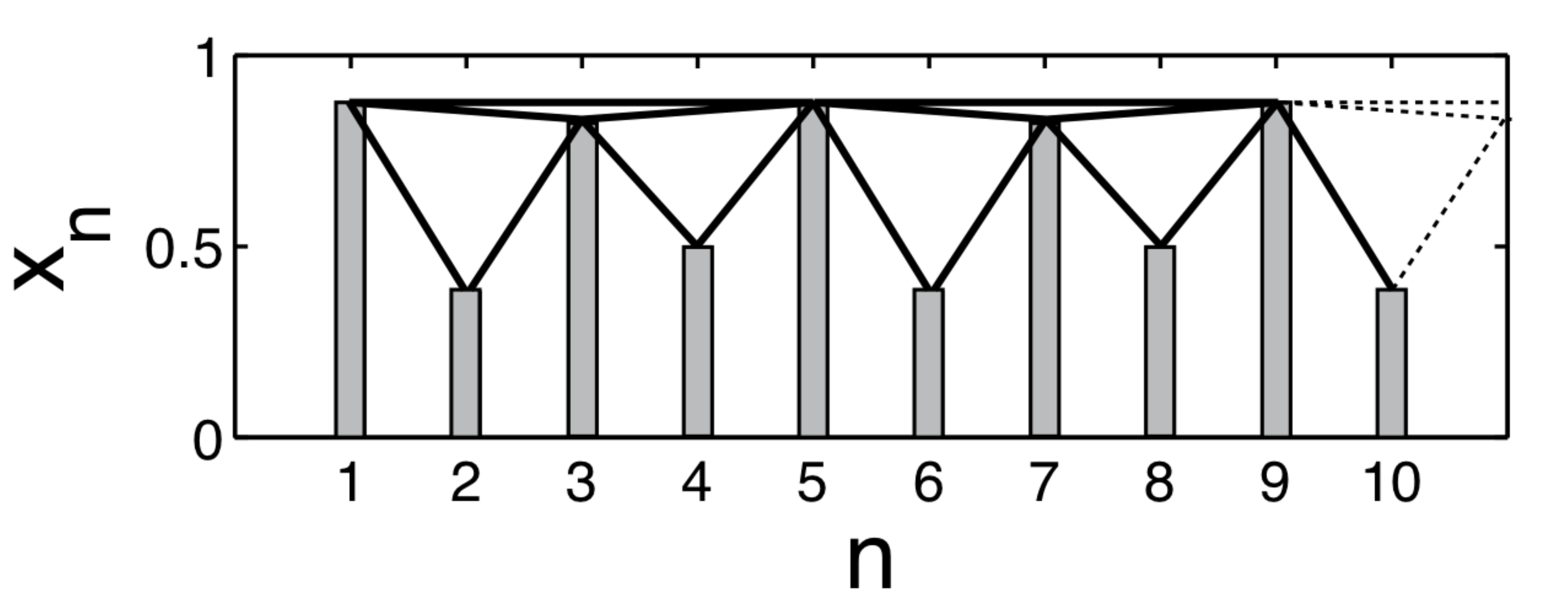}
\includegraphics[width=0.6\columnwidth]{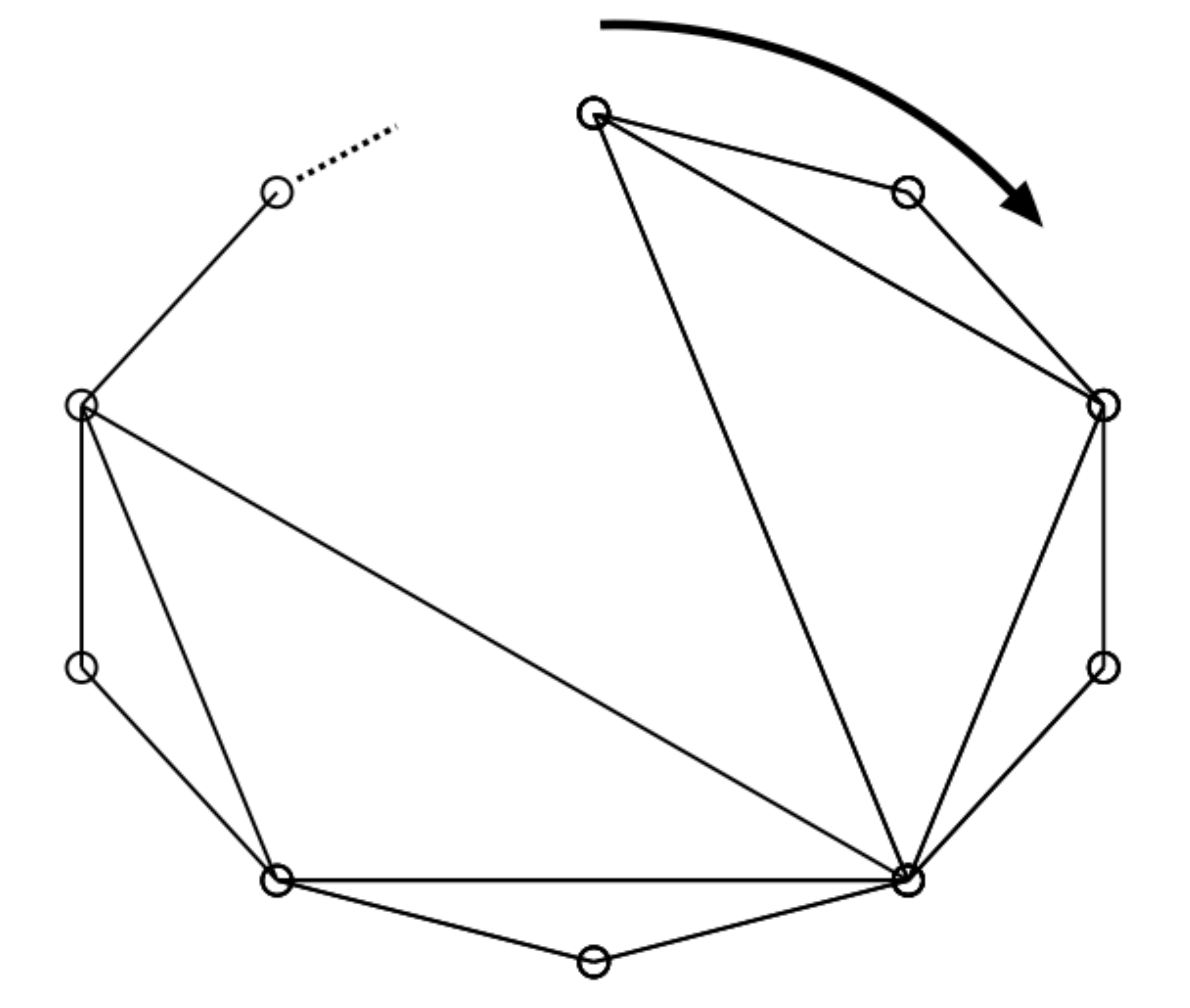}
\end{center}
\captionsetup{justification=raggedright, singlelinecheck=false}
\caption{A periodic time series (top panel) obtained from the logistic map at $\mu = 3.5$ is mapped to a network using visibility algorithm that shows repetition of motifs periodically (bottom panel).
\label{periodic}}
\end{figure}

\begin{figure}[t]
\begin{center}
\setlength{\unitlength}{0.012500in}%
\includegraphics[width=\columnwidth]{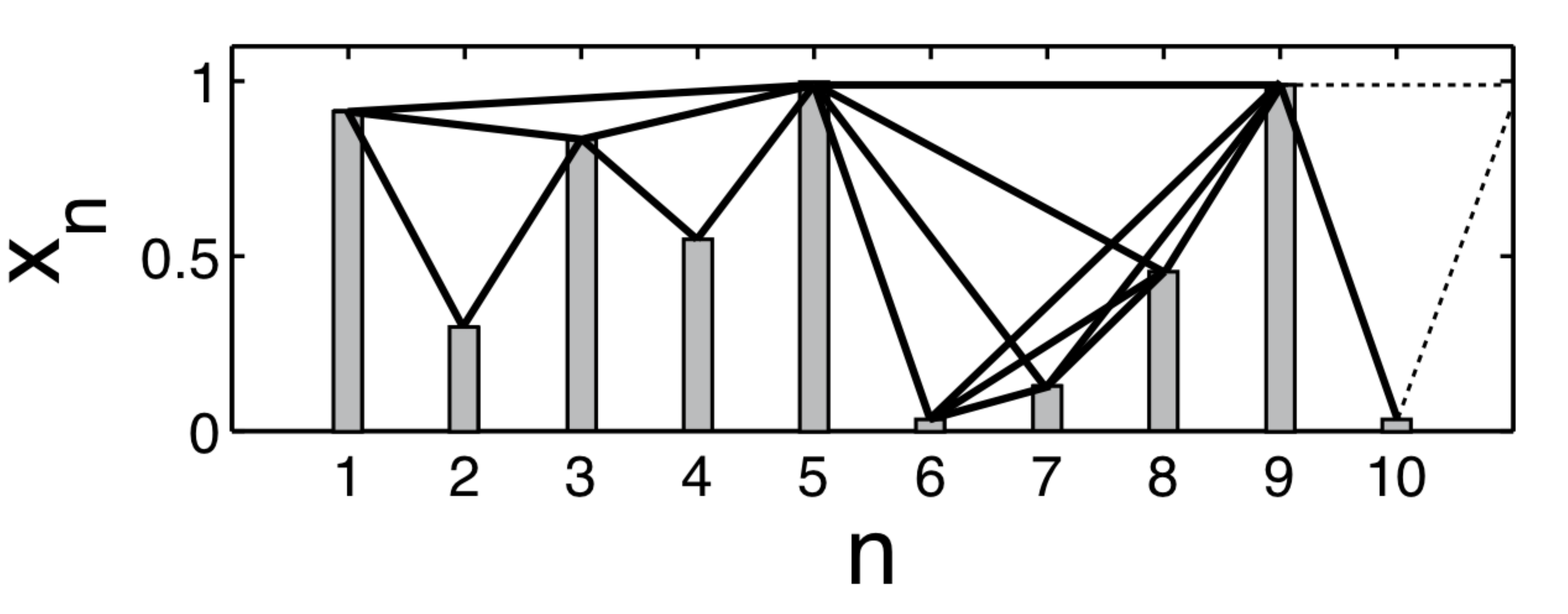}
\includegraphics[width=0.6\columnwidth]{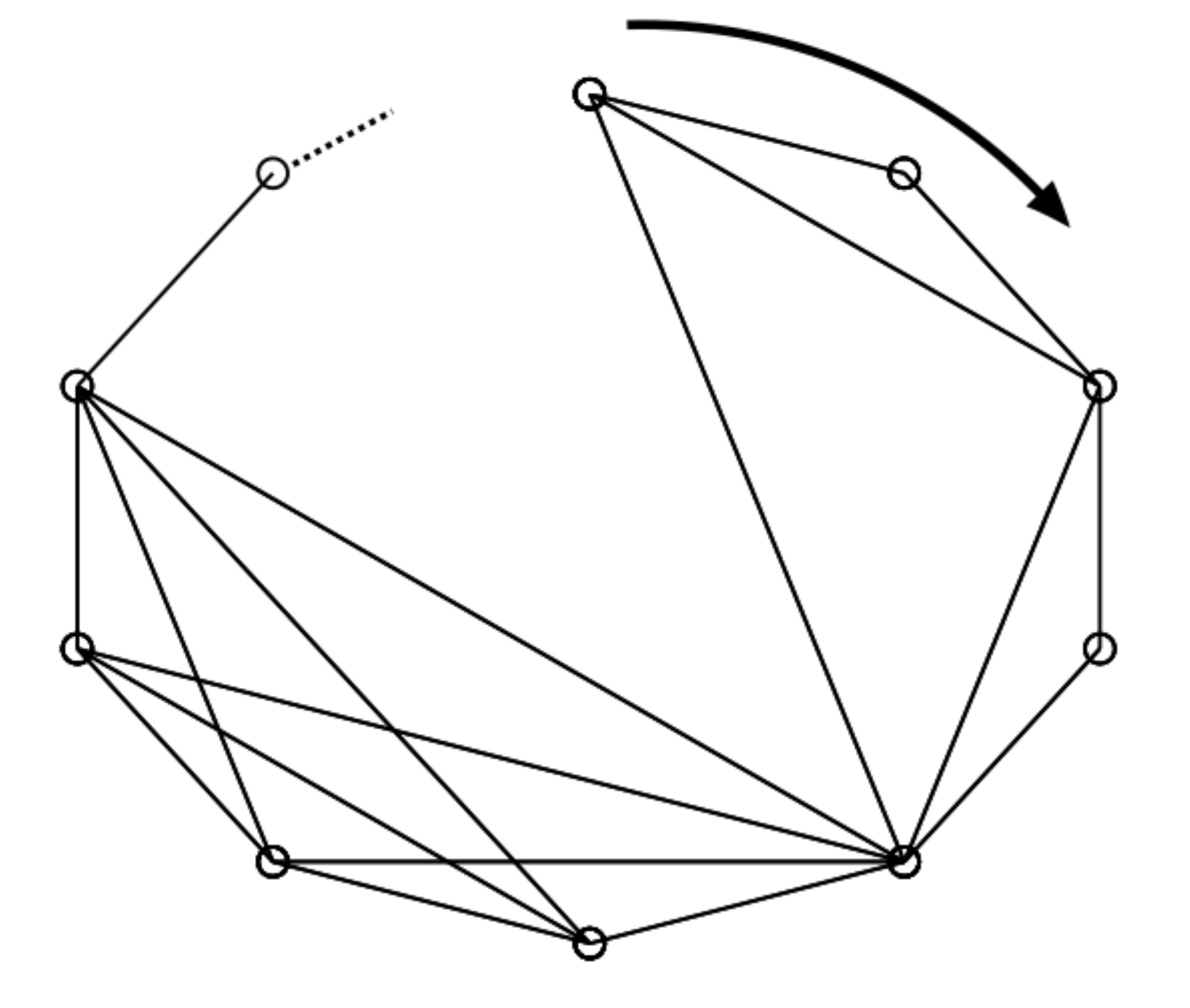}
\end{center}
\captionsetup{justification=raggedright, singlelinecheck=false}
\caption{A chaotic time series (top panel) obtained from the  logistic map at $\mu = 4.0$ and its network realization (bottom panel). This  network shows distinct features that are different from the network generated using the periodic time series in Fig.\ref{periodic}.
\label{chaotic}}
\end{figure}

In this paper, our system of interest is the logistic map, defined by $x_{n+1} = \mu x_n (1 - x_n)$, where the nonlinearity is introduced in the map by the parameter $\mu \in [0,4]$, and $x_n \in [0,1]$.  The time series of the logistic map shows periodic behaviour for $\mu = 3.5$, and chaotic behaviour for $\mu = 4.0$, as shown in the top panels of Figs. \ref{periodic} and \ref{chaotic}, respectively. We employ the visibility algorithm to convert these time series into their corresponding network representations. The network representation of the periodic time series shows repetitive motifs (Fig. \ref{periodic}), and the corresponding network for the chaotic time series shows an irregular topology (Fig. \ref{chaotic}). Similar network representations of the time series can be seen in \cite{Luque}, but have not been further analyzed by quantitative methods.

In the next section we analyze the network representations obtained from the time series at various values of $\mu$ using the methods of algebraic topology. Since the dynamical system which contributes the time series is very well understood, the TS networks studied here constitute good test beds for analyzing the effectiveness of the algebraic topology methods.
%--------------------------------------------------------------------------------------------------------------------------------------------
%--------------------------------------------------------------------------------------------------------------------------------------------
\section{The definitions of the simplicial characterizers \label{Nets}}

Here, we study the topological  structural properties of the networks generated by the visibility algorithm. The connectivity and topological properties of the network graphs reflect the connections between the dynamical states of the system in time \cite{Maletic,Kramer}. The networks so obtained are further classified using the concepts of \textit{cliques} and \textit{simplices} \cite{Bron,Andjelkovic}. We summarize these below.  

In our context,  a graph or a network represents a collection of interacting nodes interconnected by  links or edges. We define a clique to be a maximal complete subgraph, i.e. the nodes of a clique are not part of a larger complete sub-graph. Using the adjacency matrix of a the network, the Bron-Kerbosch algorithm \cite{Bron} is used to identify the cliques. The cliques are regarded as the simplicial complexes of the graph.

A simplex with $q+1$ nodes or vertices is a $q$-dimensional simplex. For instance, a $0$-simplex is an isolated point, a $1$-simplex is two vertices connected by a line segment, a $2$-simplex is a triangle of three connected nodes, a $3$-simplex is a tetrahedron with $4$ connected nodes, and so on.

Further, if two simplices have $q+1$ nodes in common, they share a $q$-face. A collection of simplices -- not just the nodes, but their shared faces as well -- forms a simplicial complex. The dimension of the simplicial complex is defined as the dimension of the largest simplex in the structure. If we can find a sequence of simplices such that each successive pair share a $q$-face, then all the simplices in this sequence are said to be $q$-connected. Simplices which are $q$-connected are also connected at all lower levels.

%--------------------------------------------------------------------------------------------------------------------------------------------
% TS AND TS-NETWORKS FOR mu=3.5 & 4.0
%--------------------------------------------------------------------------------------------------------------------------------------------
%---
%--------------------------------------------------------------------------------------------------------------------------------------------

In our study, we carry out the  structural and connectivity analysis of the TS networks obtained from the logistic map time series, using six topological characterizers, both global and local \cite{Kramer,Atkin72,Duckstein,Maletic}.
Three of these quantities are well known, and defined in most algebraic topology texts \cite{Jonsson}, and three are new and have been recently defined in the context of social and traffic networks \cite{Maletic, Andjelkovic1}.

The first characterizer is the vector  $\mathbf{Q}$, known as the first structure vector, which is a measure of the connectivity of the clique complex  at various levels. 
The $q$th component of $\mathbf{Q} = \lbrace Q_0, Q_1, \ldots Q_{q_{max}}\rbrace$ is the 
number of $q$-connected components at the $q$-th level. 
The next vector quantity, which we denote by $\mathbf{\tilde{f}}$, is defined to have the number of $q$-dimensional simplices as  its $q$-th component. The third quantity $\mathbf{N_s} = \lbrace n_0, n_1, \ldots, n_{q_{max}} \rbrace$, known as the second structure vector, has the number of simplices of dimension $q$ and higher as its $q$-th component. The fourth quantity is the third structure vector, $\mathbf{\widehat{Q}}$ which is defined in terms of the previously defined structure vectors $\mathbf{Q}$ and $\mathbf{N_s}$. Its $q$-th component, $\widehat{Q}_q$ is given by $\left(1-\frac{Q_q}{n_q}\right)$. A fifth quantity $\mathrm{dim}\, Q^{i}$, is a local quantity, which defines the topological dimension of node $i$ of the simplicial complex, given by

\begin{equation}
\mathrm{dim}\,Q^i = \sum_{q=0}^{q_{\mathrm{max}}}\,Q_q^i,
\end{equation}

where $q_{\mathrm{max}}$ is the dimension of the simplicial complex, and $Q_{k}^{i}$ is the number of different simplices of dimension $k$ in which the node $i$ participates. 

Finally, the topological entropy $\mathbf{S}$ is defined as

\begin{equation}
S_Q(q) = - \frac{\sum_i\, p_q^i\,\mathrm{log}\,p_q^i}{\mathrm{log}\,N_q}.
\end{equation}

Here, $p_q^i = {Q_q^i}/{\sum_i\, Q_q^i} $ is the probability of a particular node $i$ participating in a $q$-simplex, and $N_q = \Sigma_i \left(1 - \delta_{Q_q^i,0}\right)$ denotes the number of nodes that participate in at least one $q$-simplex.

The TS-network graphs are analyzed using these six simplicial characterizers. The calculation is illustrated for a simple example in the Appendix. The simplicial characterizers obtained for the actual TS network graphs are listed in Tables \ref{simp_charac_table_1} and \ref{simp_charac_table_2}.

%--------------------------------------------------------------------------------------------------------------------------------------------
% SIMPLICIAL CHARACS TABLES
%--------------------------------------------------------------------------------------------------------------------------------------------
\begin{table*}[!t]
\setlength{\tabcolsep}{6pt} %default is 6pt
\captionsetup{justification=raggedright, singlelinecheck=false}
\caption{Structure vectors $\mathbf{Q}$, $\mathbf{N_{s}}$, and $\mathbf{\widehat{Q}}$ for logistic map TS-networks. These TS networks are constructed from time series of length $2\,000$ (after discarding the first $5\,000$ points) using the visibility algorithm.
\label{simp_charac_table_1}}
\centering
\begin{tabular}{ *{13}{c} }
\hline
\hline
  
\multicolumn{2}{c}{}  & \multicolumn{2}{c}{$\mathbf{Periodic}$}  & \phantom{a} & \multicolumn{2}{c}{$\mathbf{Intermittent}$} & \phantom{a} & \multicolumn{1}{c}{$\mathbf{Feigenbaum}$} & \phantom{a} & \multicolumn{3}{c}{$\mathbf{Chaotic}$}\\

\cline{3-4} \cline{6-7} \cline{9-9} \cline{11-13}
&	$q$-level $/$ $\mu$	&	3.5	&	3.836	&	&	3.8284	&	3.857	&	&	3.56995	&	&	3.87	&	3.89	&	4.0	\\
& 	&	Period 4	&	Period 3	&	&	Before P3	&	&	&	&	&	Chaos 1	&	Chaos 2	&	Full chaos	\\
\\ \hline
\hspace{-0.5in} \rdelim\{{12}{20pt} \hspace{-1in}
&	0	&	1	&	1	&	&	1	&	1	&	&	1	&	&	1	&	1	&	1	\\
&	1	&	501	&	667	&	&	559	&	139	&	&	128	&	&	97	&	84	&	77	\\
&	2	&	1498	&	666	&	&	842	&	1115	&	&	1871	&	&	1384	&	1386	&	1117	\\
&	3	&		&	666	&	&	587	&	695	&	&		&	&	451	&	463	&	459	\\
$\mathbf{Q}$
&	4	&		&		&	&	10	&	48	&	&		&	&	56	&	58	&	192	\\
&	5	&		&		&	&	1	&	3	&	&		&	&	12	&	10	&	77	\\
&	6	&		&		&	&		&		&	&		&	&		&	1	&	38	\\
&	7	&		&		&	&		&		&	&		&	&		&		&	20	\\
&	8	&		&		&	&		&		&	&		&	&		&		&	15	\\
&	9	&		&		&	&		&		&	&		&	&		&		&	4	\\
\\
\hspace{-0.5in} \rdelim\{{12}{20pt} \hspace{-1in}
&	0	&	1499	&	667	&	&	853	&	1166	&	&	1872	&	&	1478	&	1489	&	1228	\\
&	1	&	1499	&	667	&	&	853	&	1166	&	&	1872	&	&	1478	&	1489	&	1228	\\
&	2	&	1498	&	666	&	&	851	&	1165	&	&	1871	&	&	1476	&	1488	&	1228	\\
&	3	&		&	666	&	&	590	&	710	&	&		&	&	474	&	483	&	509	\\
$\mathbf{N_{s}}$&
4	&		&		&	&	10	&	48	&	&		&	&	61	&	62	&	214	\\
&	5	&		&		&	&	1	&	3	&	&		&	&	12	&	10	&	92	\\
&	6	&		&		&	&		&		&	&		&	&		&	1	&	50	\\
&	7	&		&		&	&		&		&	&		&	&		&		&	25	\\
&	8	&		&		&	&		&		&	&		&	&		&		&	15	\\
&	9	&		&		&	&		&		&	&		&	&		&		&	4	\\
\\
\hspace{-0.5in} \rdelim\{{12}{20pt} \hspace{-1in}
&	0	&	0.9993	&	0.9985	&	&	0.9988	&	0.9991	&	&	0.9995	&	&	0.9993	&	0.9993	&	0.9992	\\
&	1	&	0.6658	&	0	&	&	0.3447	&	0.8808	&	&	0.9316	&	&	0.9344	&	0.9436	&	0.9373	\\
&	2	&	0	&	0	&	&	0.0106	&	0.0429	&	&	0	&	&	0.0623	&	0.0685	&	0.0904	\\
&	3	&		&	0	&	&	0.0051	&	0.0211	&	&		&	&	0.0485	&	0.0414	&	0.0982	\\
$\mathbf{\widehat{Q}}$
&	4	&		&		&	&	0	&	0	&	&		&	&	0.0820	&	0.0645	&	0.1028	\\
&	5	&		&		&	&	0	&	0	&	&		&	&	0	&	0	&	0.1630	\\
&	6	&		&		&	&		&		&	&		&	&		&	0	&	0.24	\\
&	7	&		&		&	&		&		&	&		&	&		&		&	0.2	\\
&	8	&		&		&	&		&		&	&		&	&		&		&	0	\\
&	9	&		&		&	&		&		&	&		&	&		&		&	0	\\
\hline
\end{tabular}
\end{table*}
%--------------------------------------------------------------------------------------------------------------------------------------------

\begin{table*}[!t]
\setlength{\tabcolsep}{6pt} %default is 6pt
\captionsetup{justification=raggedright, singlelinecheck=false}
\caption{The topological response function $\mathbf{\tilde{f}}$ and the topological entropy $\mathbf{S}$ for the logistic map TS-networks. These TS-networks are constructed from a time series of length $2\,000$ (after discarding the first $5\,000$ points) using the visibility algorithm.
\label{simp_charac_table_2}}
\centering
\begin{tabular}{ *{13}{c} }
\hline
\hline
  
\multicolumn{2}{c}{}  & \multicolumn{2}{c}{$\mathbf{Periodic}$}  & \phantom{a} & \multicolumn{2}{c}{$\mathbf{Intermittent}$} & \phantom{a} & \multicolumn{1}{c}{$\mathbf{Feigenbaum}$} & \phantom{a} & \multicolumn{3}{c}{$\mathbf{Chaotic}$}\\

\cline{3-4} \cline{6-7} \cline{9-9} \cline{11-13}
&	$q$-level $/$ $\mu$	&	3.5	&	3.836	&	&	3.8284	&	3.857	&	&	3.56995	&	&	3.87	&	3.89	&	4.0	\\
& 	&	Period 4	&	Period 3	&	&	Before P3	&	&	&	&	&	Chaos 1	&	Chaos 2	&	Full chaos	\\
\\ \hline
\hspace{-0.5in} \rdelim\{{12}{20pt} \hspace{-1in}
&	0	&	0	&	0	&	&	0	&	0	&	&	0	&	&	0	&	0	&	0	\\
&	1	&	1	&	1	&	&	2	&	1	&	&	1	&	&	2	&	1	&	0	\\
&	2	&	1498	&	0	&	&	261	&	455	&	&	1871	&	&	1002	&	1005	&	719	\\
&	3	&		&	666	&	&	580	&	662	&	&		&	&	413	&	421	&	295	\\
$\mathbf{\tilde{f}}$
&	4	&		&		&	&	9	&	45	&	&		&	&	49	&	52	&	122	\\
&	5	&		&		&	&	1	&	3	&	&		&	&	12	&	9	&	42	\\
&	6	&		&		&	&		&		&	&		&	&		&	1	&	25	\\
&	7	&		&		&	&		&		&	&		&	&		&		&	10	\\
&	8	&		&		&	&		&		&	&		&	&		&		&	11	\\
&	9	&		&		&	&		&		&	&		&	&		&		&	4	\\
\\
\hspace{-0.5in} \rdelim\{{12}{20pt} \hspace{-1in}
&	0	&	0	&	0	&	&	0	&	0	&	&	0	&	&	0	&	0	&	0	\\
&	1	&	1	&	1	&	&	0.9464	&	1	&	&	1	&	&	1	&	1	&	0	\\
&	2	&	0.9774	&	0	&	&	0.9697	&	0.9752	&	&	0.9620	&	&	0.9779	&	0.9765	&	0.9742	\\
&	3	&		&	0.9923	&	&	0.9919	&	0.9838	&	&		&	&	0.9783	&	0.9766	&	0.9786	\\
$\mathbf{S}$
&	4	&		&		&	&	0.9763	&	0.9806	&	&		&	&	0.9824	&	0.9848	&	0.9869	\\
&	5	&		&		&	&	1	&	1	&	&		&	&	0.9707	&	0.9731	&	0.9912	\\
&	6	&		&		&	&		&		&	&		&	&		&	1	&	0.9930	\\
&	7	&		&		&	&		&		&	&		&	&		&		&	0.9758	\\
&	8	&		&		&	&		&		&	&		&	&		&		&	0.9938	\\
&	9	&		&		&	&		&		&	&		&	&		&		&	1	\\
\hline
\end{tabular}
\end{table*}
%--------------------------------------------------------------------------------------------------------------------------------------------
% MAXDIMQ TABLE
%--------------------------------------------------------------------------------------------------------------------------------------------
\begin{table}
\captionsetup{justification=raggedright}
\caption{The maximum value of the topological dimension of all nodes in the TS-network of the logistic map at the parameter values indicated in the table. The time series considered is of length 2\,000.}
\label{maxDimQ_table}
\setlength{\tabcolsep}{12pt}
\begin{tabular}{ *{3}{c} }
\hline
\hline
$\mu$ 	& 	Nature of Orbit 	& 	max(dim $Q^{i}$)	\\
\hline
3.5	& 	Period 4 	& 	4	\\
3.836	& 	Period 3 	& 	2	\\
\\
3.56995	& 	Feigenbaum point 	& 	8	\\
\\
3.8284	& 	Intermittency before P3 	& 	8	\\
3.857	& 	Intermittency 	& 	12	\\
\\
3.87	& 	Chaos 1	& 	13	\\
3.89	& 	Chaos 2	& 	13	\\
4	& 	Full chaos 	& 	12	\\
\hline
\end{tabular}
\end{table}
%--------------------------------------------------------------------------------------------------------------------------------------------
%--------------------------------------------------------------------------------------------------------------------------------------------
\section{Results for the logistic map time series \label{results}}

In this section, our objective is to investigate the connection between the topological structure arising out of the TS network, and the underlying dynamics of the evolving system, the logistic map.
In other words, we are interested in examining the signature of specific dynamical behaviors of the base system corresponding to the topological structure of the respective networks.
The time series of the logistic map has been obtained at eight distinct values of parameters, where distinct classes of dynamical behavior are seen. These parameter values have been indicated in the bifurcation diagram in Fig.\,\ref{logisticMap} and include representative samples, from the periodic, intermittent, and chaotic regimes of the logistic map, and also at the edge of chaos. The six simplicial  characterizers defined  above, have been calculated for the TS-networks obtained using the time series data for  $t=2\,000$ time steps, after discarding the initial $5\,000$ transients. Therefore, the TS networks have $N=2\,000$ nodes. The visibility condition (given in Equation \ref{visibility}) has been implemented with a tolerance of $\epsilon=10^{-4}$.

The dynamical regimes are:

\begin{enumerate}

\item The periodic regime: This was studied at two parameter values, viz. 
$\mu = 3.5$ (period-4 behaviour), and
$\mu = 3. 836$ (period-3 window). 

\item The edge of chaos: $\mu = 3.56995$ (Feigenbaum point, period-2 cascade ends here)

\item Intermittency :  For this we studied the values
$\mu = 3.8284$ (onset of crisis induced intermittency), and
$\mu = 3.857$ (post-crisis induced intermittency). 

\item  The chaotic dynamical regime are
$\mu = 3.87$ (chaos),
$\mu = 3.89$ (chaos), and
$\mu = 4.0$ (fully developed chaos).

\end{enumerate}
 
As mentioned above the time series data at these parameter values is transformed to networks using the visibility algorithm, and the clique structure of the resulting network is extracted using the Bron-Kerbosch algorithm \cite{Bron}. Further, the six topological quantities are calculated for these networks. The results of the analysis are presented in the Tables \ref{simp_charac_table_1}, \ref{simp_charac_table_2}, and \ref{maxDimQ_table}. Two central aspects of the network characterizers among the six quantities are: topological structure and topological connectivity.

%--------------------------------------------------------------------------------------------------------------------------------------------
% 10k TS SIMPLICIAL CHARACS PLOTS (Q, Ns, Qhat)
%--------------------------------------------------------------------------------------------------------------------------------------------
\begin{figure*}
\begin{tabular}{c c}
	\subfloat[\label{10k_simp_characs_1}]{\includegraphics[width=\columnwidth]{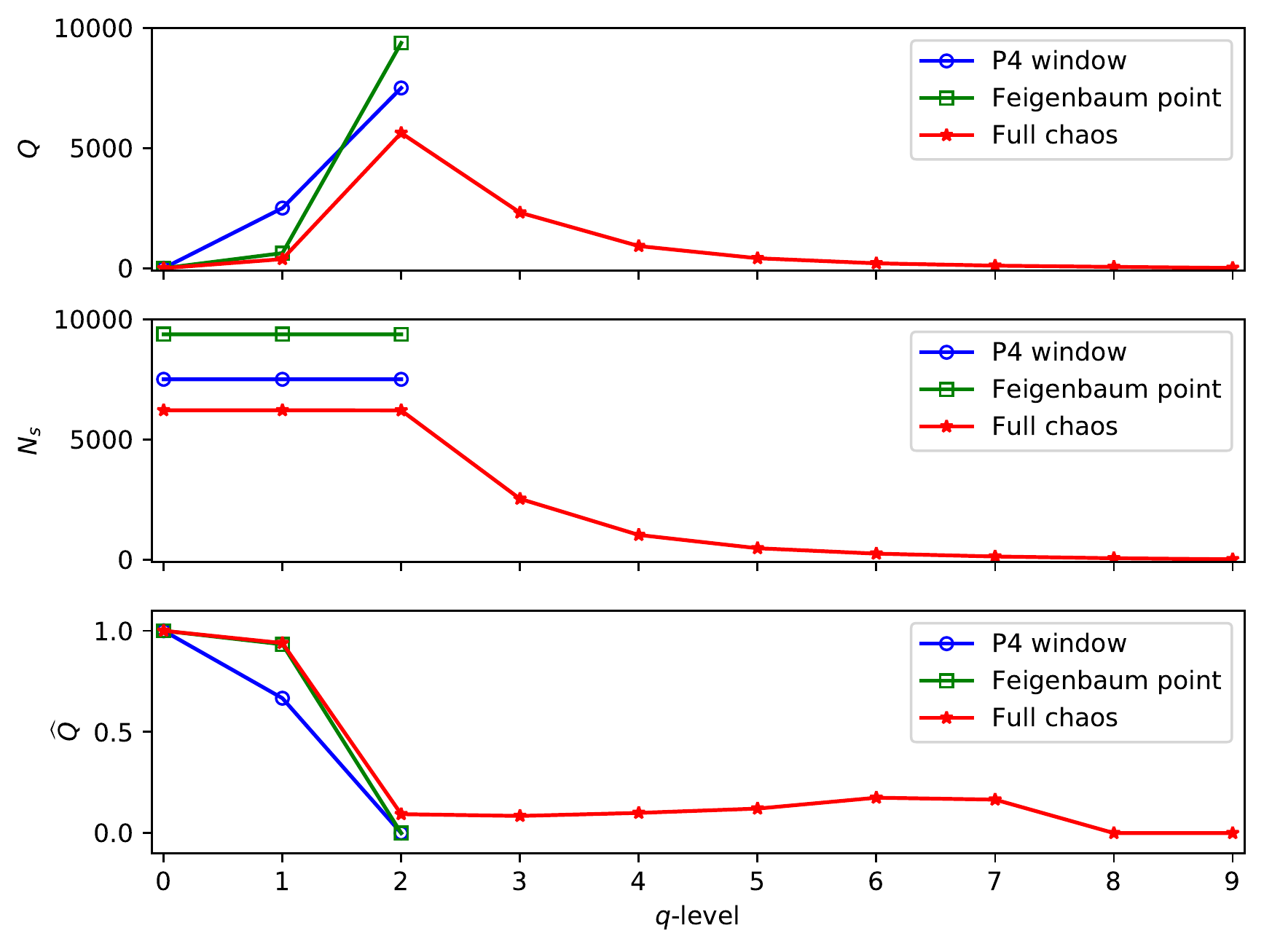}}
	&
	\subfloat[\label{10k_simp_characs_4}]{\includegraphics[width=\columnwidth]{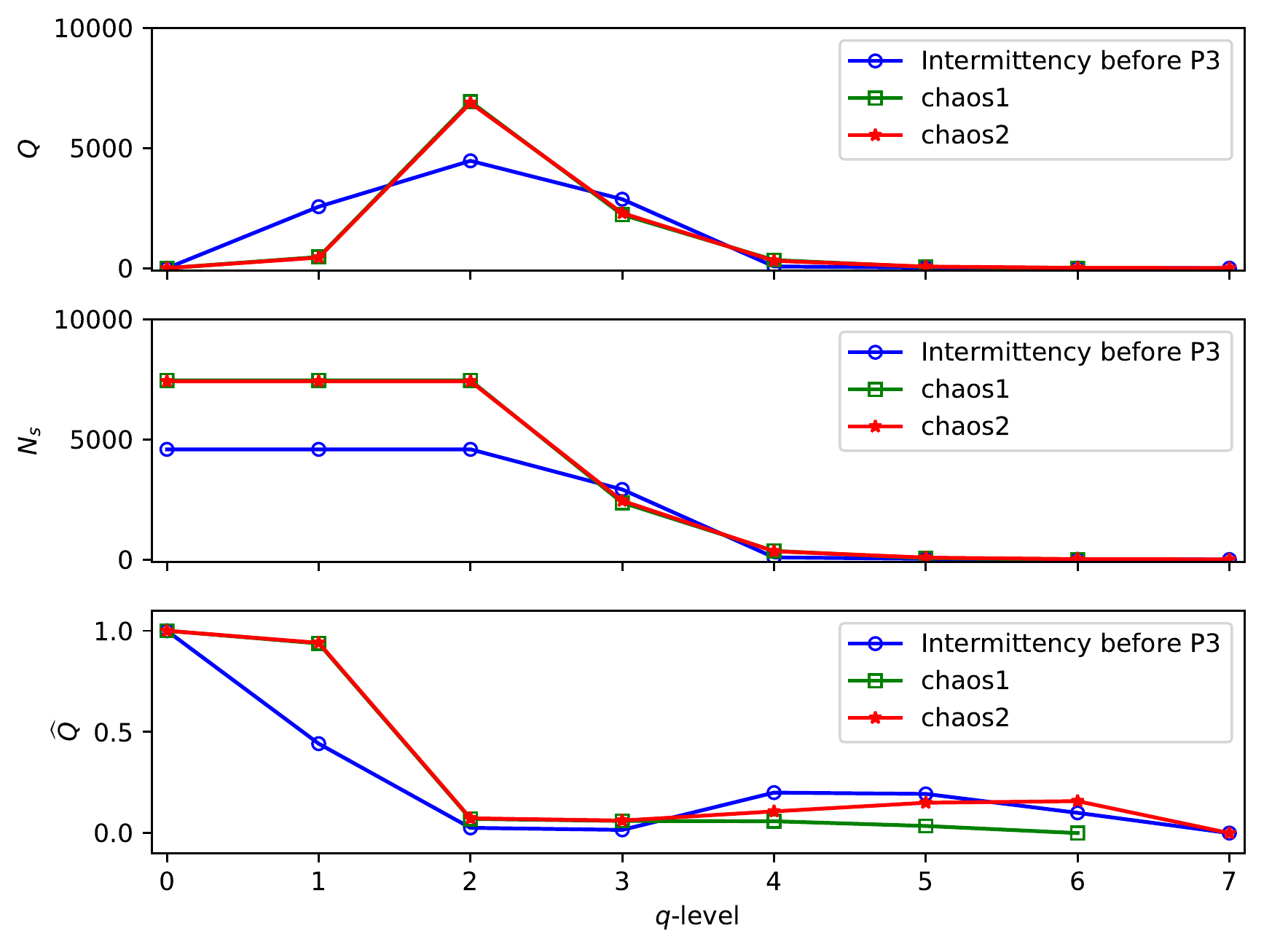}}
\end{tabular}
\captionsetup{justification=raggedright, singlelinecheck=false}
\caption{ (Color online) Structure vectors $\mathbf{Q}$, $\mathbf{N_{s}}$, and $\mathbf{\widehat{Q}}$ for logistic map TS-networks. These TS networks are constructed from time series of length $10\,000$ (after discarding the first $5\,000$ steps) using the visibility algorithm.
Fig.\,\ref{10k_simp_characs_1} is for parameter values $\mu = 3.5$ (P4 window), $\mu = 3.56995$ (Feigenbaum point), and $\mu = 4$ (Full chaos).
Fig.\,\ref{10k_simp_characs_4} is for parameter values $\mu = 3.8284$ (Intermittency before P3), $\mu = 3.87$ (chaos 1), and $\mu = 3.89$ (chaos 2).
\label{10k_simps_1}}
\end{figure*}
%--------------------------------------------------------------------------------------------------------------------------------------------

%--------------------------------------------------------------------------------------------------------------------------------------------
% 10k TS SIMPLICIAL CHARACS PLOTS (f, S)
%--------------------------------------------------------------------------------------------------------------------------------------------
\begin{figure*}[!t]
\begin{tabular}{c c}
	\subfloat[\label{10k_simp_characs_2}]{\includegraphics[width=\columnwidth]{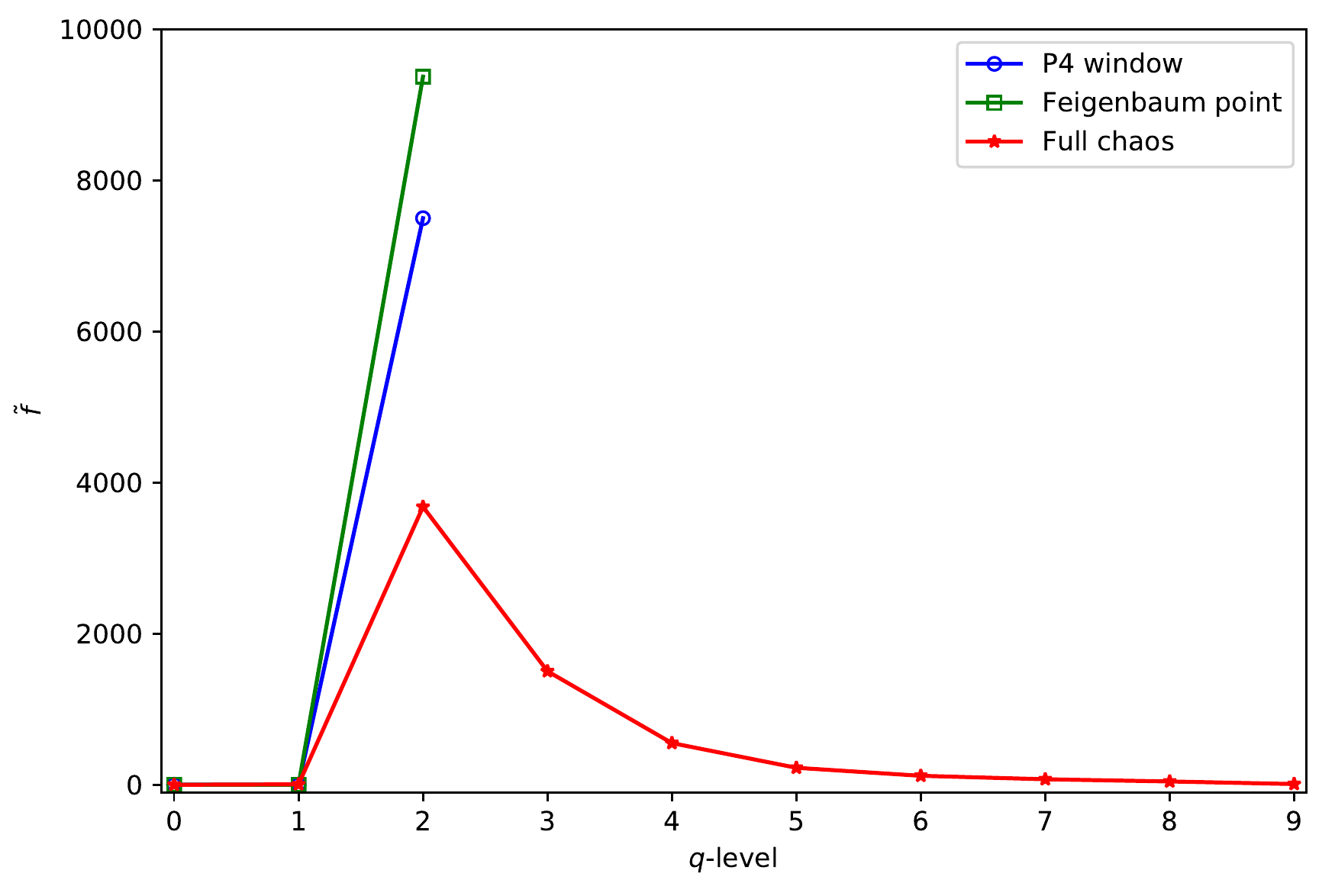}}
	&
	\subfloat[\label{10k_simp_characs_5}]{\includegraphics[width=\columnwidth]{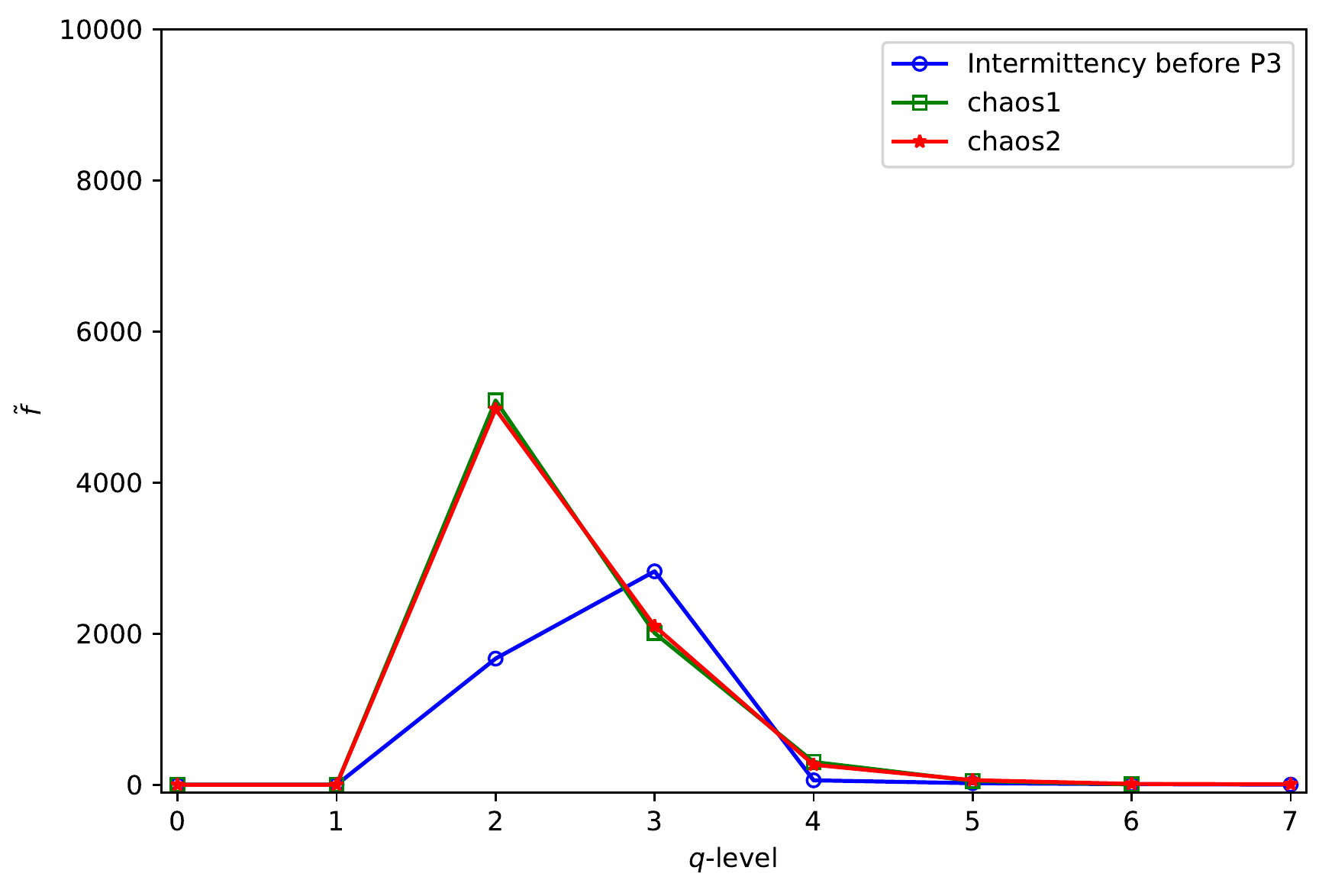}}
	\\
	\subfloat[\label{10k_simp_characs_3}]{\includegraphics[width=\columnwidth]{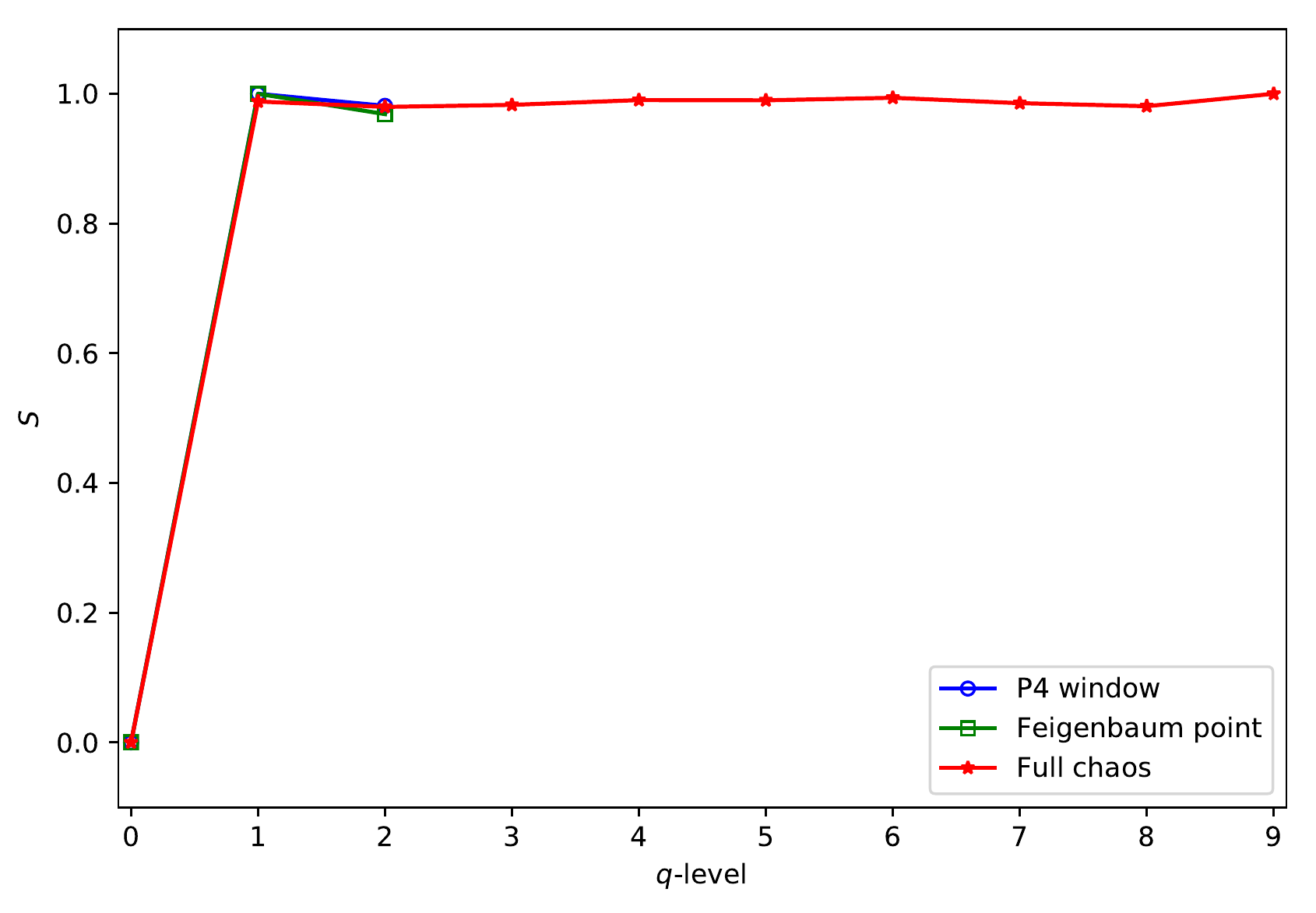}}
	&
	\subfloat[\label{10k_simp_characs_6}]{\includegraphics[width=\columnwidth]{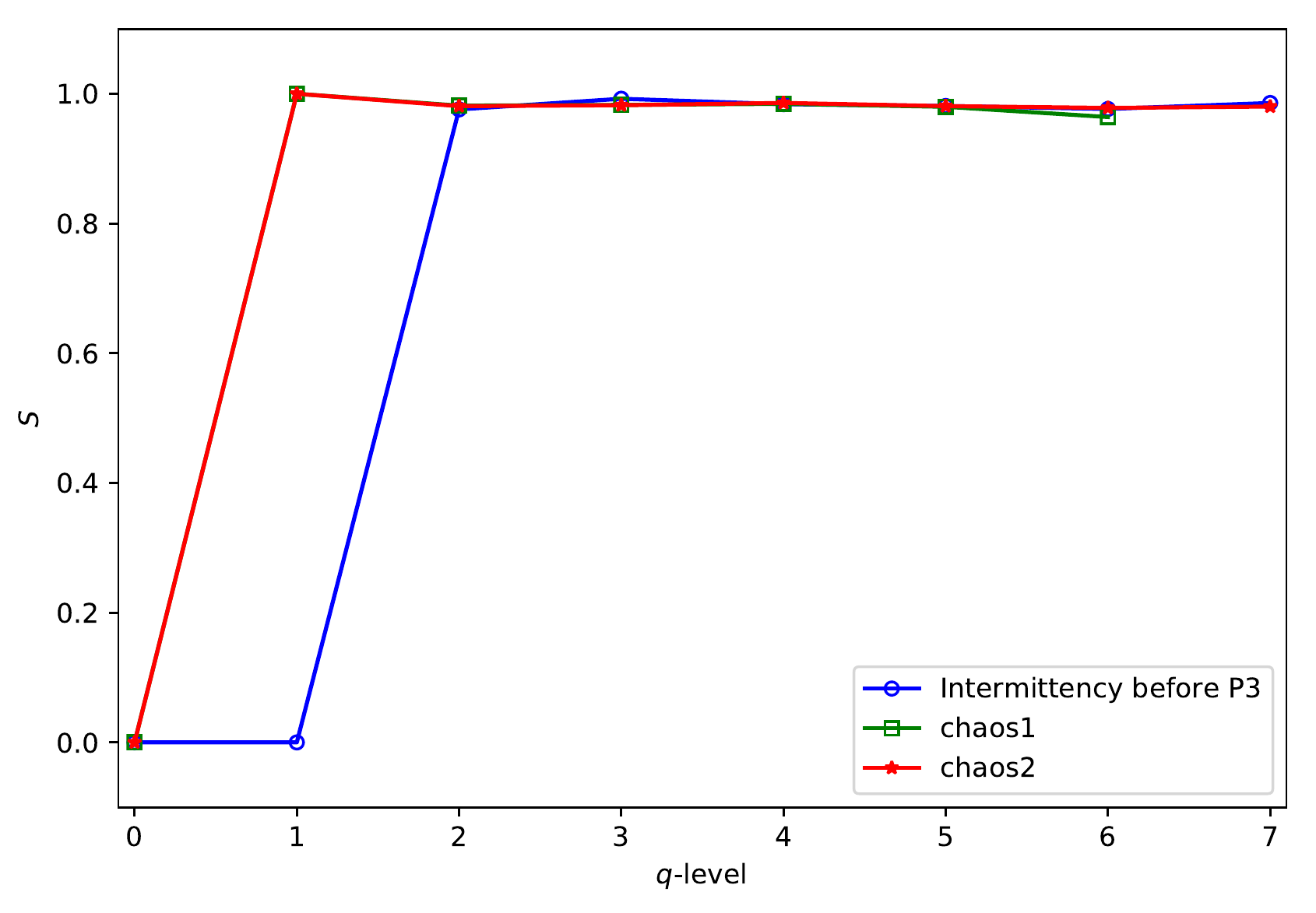}}
\end{tabular}
\captionsetup{justification=raggedright, singlelinecheck=false}
\caption{(Color online) Topological response function $\mathbf{\tilde{f}}$ and topological entropy $\mathbf{S}$ for logistic map TS-networks. These TS networks are constructed from time series of length $10\,000$ ( the first $5\,000$ steps) using the visibility algorithm.
Figs.\,\ref{10k_simp_characs_2} and \ref{10k_simp_characs_3} are for parameter values $\mu = 3.5$ (P4 window), $\mu = 3.56995$ (Feigenbaum point), and $\mu = 4$ (Full chaos).
Figs.\,\ref{10k_simp_characs_5} and \ref{10k_simp_characs_6} are for parameter values $\mu = 3.8284$ (Intermittency before P3), $\mu = 3.87$ (chaos 1), and $\mu = 3.89$ (chaos 2).
\label{10k_simps_2}}
\end{figure*}
%--------------------------------------------------------------------------------------------------------------------------------------------

We also plot the corresponding quantities for a network of $10\,000$ nodes (i.e.\,a time series with 10\,000 points) in Figs.\,\ref{10k_simps_1} and \ref{10k_simps_2}. We note that similar trends are seen in the Tables (2\,000 node TS network) and Figures (10\,000 node TS network).

\begin{enumerate}

\item

The topological connectivity between the simplices at each topology level is identified by the simplicial characteriser $\mathbf{Q}$. The vector $\mathbf{Q}$ measures the number of connected components of the network at each topology level -- for $q=0$ (at least one vertex in common), $q=1$ (at least two vertices in common) and $q=2$ (at least three vertices in common), etc.  We observe that at the lowest topology level ($q=0$) the components of $\mathbf{Q}$ have a value of $1$ confirming that there is no isolated node in the network, for any of the parameter values, in any of the dynamical regimes.  At the higher levels, results in different dynamical regimes differ very sharply. We plot $ \mathbf{Q}$ versus $q$ in Fig.\,\ref{10k_simp_characs_1} (top panel) for the periodic and chaotic regimes. It is clear from both the tables and the graph, that the periodic regimes contribute only for the first three/four levels, whereas the chaotic regimes have contributions at much higher levels (upto level $9$ here for both the $2000$ and $10000$ point time series, with the visibility condition implemented to a tolerance of $\epsilon=10^{-4}$).

We also plot the simplicial characteriser $\mathbf{Q}$ for the intermittent value ($\mu=3.8284$) and the chaotic values $\mu=3.87, 3.89$ in Fig. {\ref{10k_simp_characs_4}} .
The chaotic value peaks much more sharply at the $q=2$ level, before it falls off, as compared to the intermittent case which both rises and falls more gently, i.e. the number of connected components at each level changes  much more gradually. We note that at $\mu=3.87$ we see contributions upto the $q=6$ level and unto the $q=7$ level for $\mu=3.89$, whereas at $\mu=4.0$ the contributions go to the $q=9$ level.

\item 

The second structure vector  $\mathbf{N_s}$ is a running index which counts the number of connected components at level $q$ and above, i.e.\,it is a cumulative index for $\mathbf{Q}$. It therefore contains the same information as seen in $\mathbf{Q}$ at the $q=9$ level, as can be seen from Fig.\,\ref{10k_simp_characs_1} (middle panel), for the chaotic regimes. Again the periodic and chaotic regimes show completely distinct behavior, with contributions in the periodic regime, and the Feigenbaum point being confined to the first three $q$-levels, whereas the the $\mu=4$ value sees contributions unto the $q=9$ level.

The intermittent case $\mu=3.8284)$ and the chaotic value $\mu=3.89$ again reflects the difference seen in the case of the first structure vector (Fig.\,\ref{10k_simp_characs_4} middle panel).

\item

The components of the third structure vector are defined in terms of the extent to which the ratio $\frac{Q_q}{n_q}$ differs from $1$. This quantity thus lies between zero and one.
These are plotted in the Fig.\,{\ref{10k_simp_characs_1}} (bottom panel), for the period 4 case, the Feigenbaum point and the chaotic case. Here again, there is a sharp difference between the periodic and chaotic cases, and the intermittency at $\mu=3.8284)$ and the chaotic value $\mu=3.87$ and $\mu=3.89$ (Fig.\,\ref{10k_simp_characs_4} bottom panel).

\item

The $\mathbf{\tilde{f}}$ vector identifies the topological structure of the network. This quantity counts the number of simplices at each topological level. This quantity functions like a response function. We see that the periodic regimes behave quite distinctly from the chaotic regimes. (See Fig.\,\ref{10k_simp_characs_2}). In the periodic regimes, the response function $\mathbf{\tilde{f}}$ rises sharply so that most of the simplices are at the topmost level, whereas in the chaotic regimes, the response functions peak sharply at the third level ($q=2$) and then fall off gradually, so that the plot has a long tail, extending unto $q=9$. The differences between the intermittent value at $\mu=3.8284)$ and the chaotic values $\mu=3.87, 3.89$ also show up clearly in the Fig.\,\ref{10k_simp_characs_5} with a clear shift in the level at which the peak occurs, and also in the height of the  peak.  The highest contributions are now at the $q=6$ and $q=7$ levels only, as compared to the $\mu=4.0$ case which has contributions till the $q=9$ levels. Thus the case of fully developed chaos is clearly differentiated from the others, by its long tail.    

\item

The topological entropy $\mathbf{S}(q)$ is a measure of the complexity of the network as well. This is plotted in Fig.\,\ref{10k_simp_characs_3}  for the period 4 case, the Feigenbaum point, and fully developed chaos at $\mu=4.0$; and in Fig.\,\ref{10k_simp_characs_6} for the intermittent case $\mu=3.8284$ and the chaotic cases ($\mu=3.87, 3.89$). The entropies of the periodic states and the edge of chaos contribute upto the $q=2$ levels, whereas the fully developed chaos state shows contributions till the $q=9$ level. 
The intermittent cases and the chaotic cases show small fluctuations relative to each other at the different levels, which go up to the $q=6,7$ levels.

\item

The quantity $\mathrm{max (dim)}\, Q^{i}$ gives the maximum value of the topological dimension of all the nodes in the network. (Refer to Table \ref{maxDimQ_table}.) This picks the changes in the dynamic regimes most strongly.

We see a clear distinction between the periodic states, and the intermittent and chaotic states in this case. For the periodic states, the node in the network which participates in the most number of simplices, participates in very few simplices, the values being $4 $ and $ 2$, for the period 3 and period 4 values. At the edge of chaos, viz. $\mu=3.56995$ there is at least one node which participate in $8$ simplicial complexes. Higher values, viz. $8$ and $13$ are seen in chaotic and intermittent regimes. At $\mu=4.0$, viz. fully developed chaos, there is a node which participates in as many as $12$ simplicial complexes. Thus there are many more interconnections in the chaotic regimes, compared to the periodic ones.
\end{enumerate}

%--------------------------------------------------------
The values of the topological characterisers in Tables \ref{simp_charac_table_1} and \ref{simp_charac_table_2} are for a $2\,000$ node network obtained out of a time  series evolving from a single initial condition.
The same qualitative features are observed for average simplicial characterisers averaged over $100$ initial conditions, as well as for longer time series of $10\,000$ points.
%--------------------------------------------------------------------------------------------------------------------------------------------
%--------------------------------------------------------------------------------------------------------------------------------------------
\subsection{The $\mathrm{max (dim)}\, Q^{i}$ and the Lyapunov exponent}

\begin{figure*}[!t]
\begin{tabular}{c}
	\subfloat[\label{MaxDimQ_plot}]{\includegraphics[width=2\columnwidth]{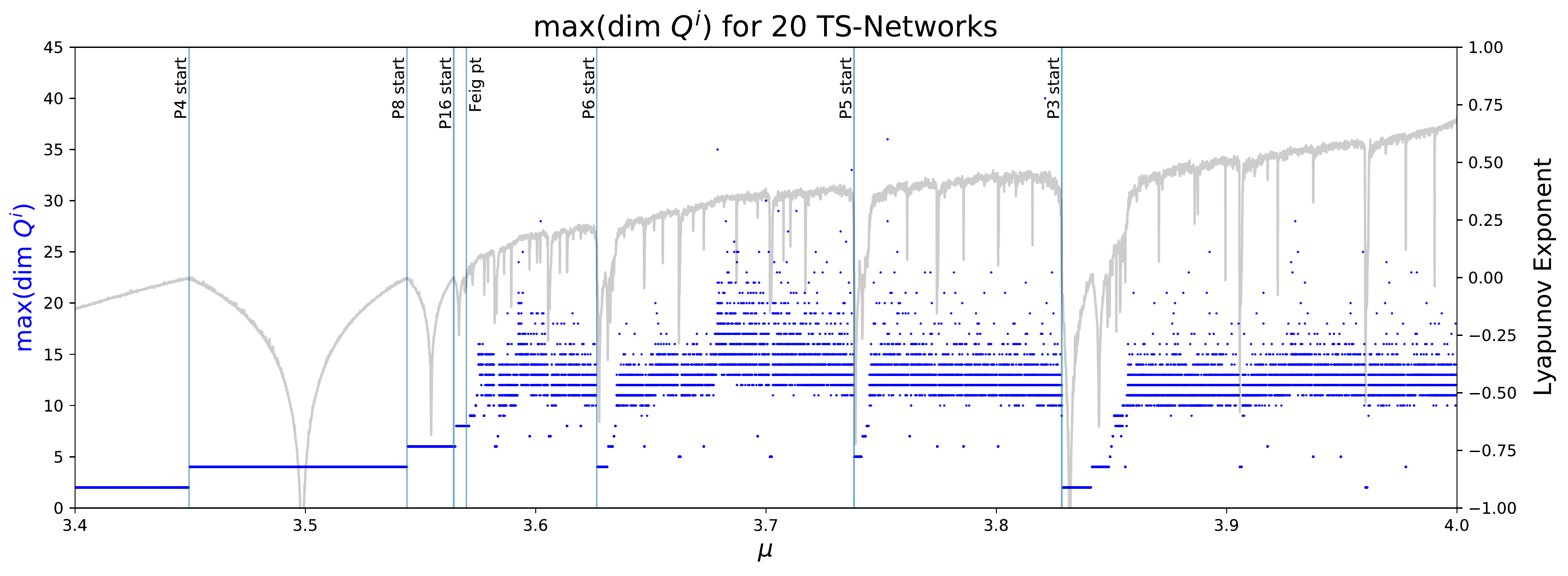}}
	\\
	\subfloat[\label{Avg_MaxDimQ_plot}]{\includegraphics[width=2\columnwidth]{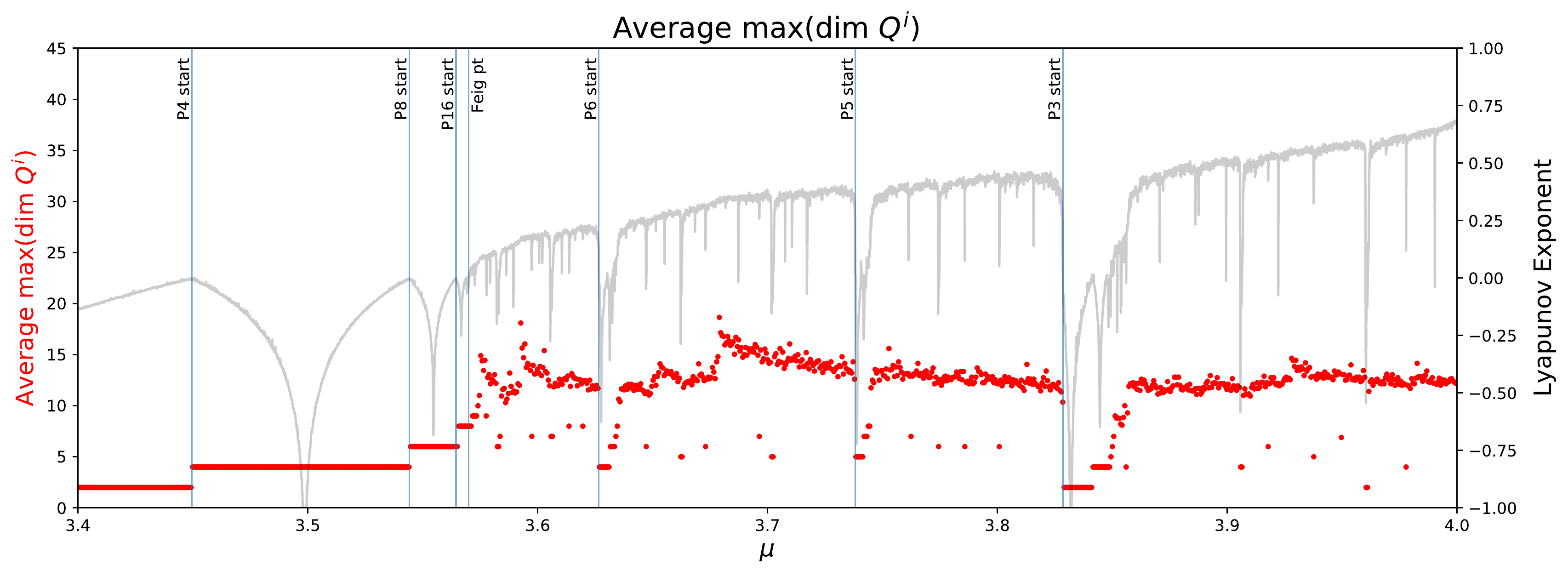}}
\end{tabular}
\captionsetup{justification=raggedright, singlelinecheck=false}
\caption{Maximum topological dimension for logistic map TS networks as a function of parameter $\mu$.
\label{MaxDimQ_Plot}}
\end{figure*}

The $\mathrm{max (dim)}\, Q^{i}$, i.e the dimension of the node that participates in the largest number of simplices of any dimension, is a measure of the complexity of the correlations in the time series at that value of the parameter $\mu$. It is interesting to compare its behaviour with the Lyapunov exponent, which is a measure of the chaoticity, as encoded by the rate of divergence of two neighbouring trajectories at the given value of the parameter. Figure \,\ref{MaxDimQ_Plot} plots the $\mathrm{max (dim)}\, Q^{i}$ as a function of the parameter $\mu$ for the parameter range $0 \leq \mu \leq 4.0$, as well as the Lyapunov exponent versus $\mu$ for the same range. 

The plot for $\mathrm{max (dim)}\, Q^{i}$ is plotted for $20$ distinct initial conditions for each $\mu$ value. The mean value of the $\mathrm{max (dim)}\, Q^{i}$ is plotted on the same graph.

\begin{figure*}[!t]
\begin{tabular}{c}
	\subfloat[\label{MaxDimQ_1}]{\includegraphics[width=2\columnwidth]{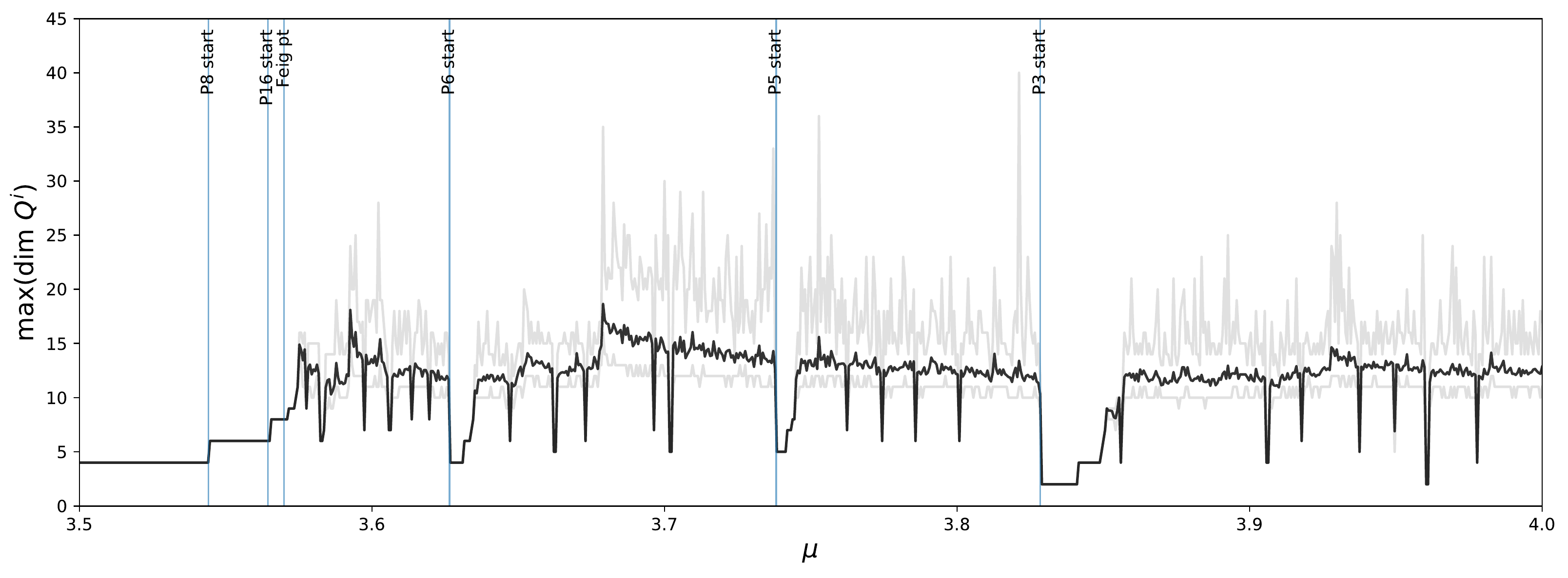}}
	\\
	\subfloat[\label{MaxDimQ_2}]{\includegraphics[width=2\columnwidth]{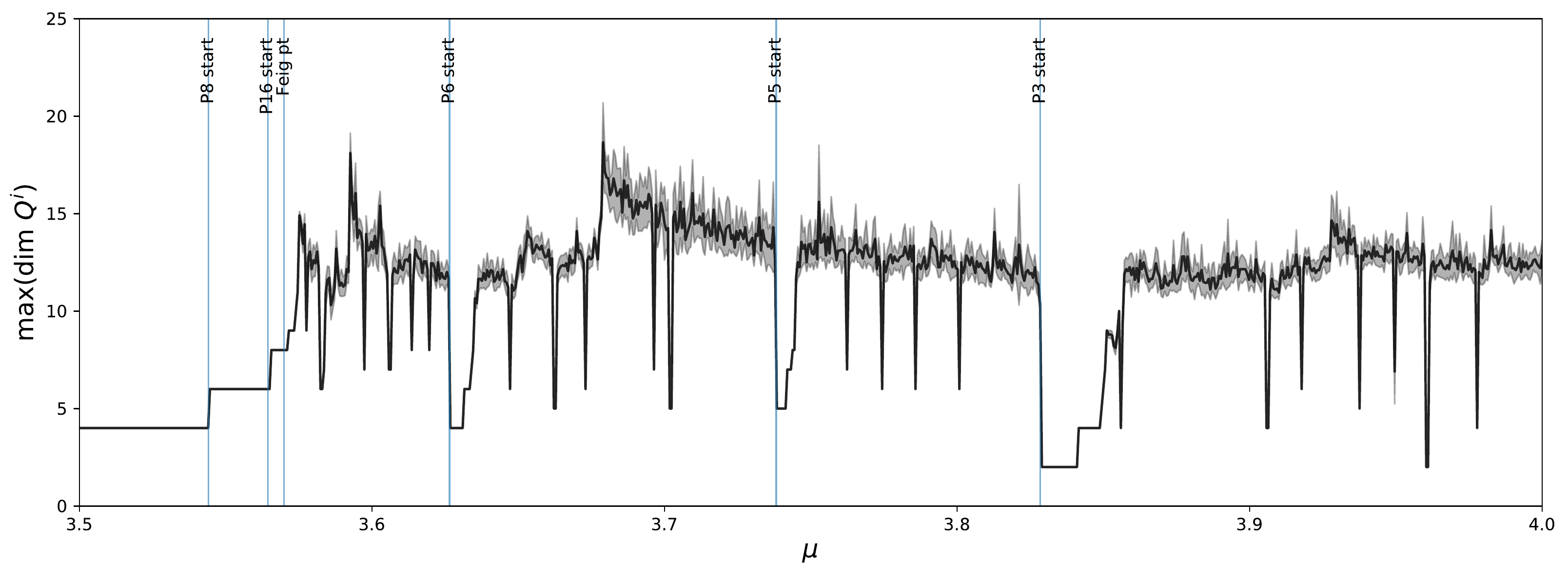}}
\end{tabular}
\captionsetup{justification=raggedright, singlelinecheck=false}
\caption{Maximum topological dimension for logistic map TS networks as a function of parameter $\mu$.
(a) shows the average value of max(dim $Q^{i}$) over 20 instances of TS networks for each $\mu$ (in black), along with the minimum and maximum value (in gray). (b) shows the average value of max(dim $Q^{i}$) and its standard deviation.
\label{MaxDimQ_Plot_2}}
\end{figure*}

It is instructive  to compare the behaviour of the $\mathrm{max (dim)}\, Q^{i}$ with that of the Lyapunov exponent in different parameter regimes. In the period doubling regime, we first note that the $\mathrm{max (dim)}\, Q^{i}$ is a constant across the window of each period and jumps at each period doubling bifurcation, indicating the change in the network connectivity that reflects each period. It is also the same for all the initial conditions, in the stable regime of the period. We note that the cascade accumulates at the Feigenbaum point at $\mu=3.56995\ldots$. Beyond this point, the fact that the trajectories have now crossed into the chaotic regime is reflected by a jump to a higher value of $\mathrm{max (dim)}\, Q^{i}$. The value of $\mathrm{max (dim)}\, Q^{i}$ is now much more sensitive to initial conditions, as is expected in the chaotic regime, and the average value therefore fluctuates much more. The existence of periodic windows in the chaotic regime is signalled by a corresponding drop in the values of $\mathrm{max (dim)}\, Q^{i}$, at the appropriate values of $\mu$.  We note that the average value of $\mathrm{max (dim)}\, Q^{i}$ fluctuates in a narrow band in the entire chaotic regime, similar to the behaviour of the Lyapunov exponent. We also note that increased sensitivity to initial conditions is seen at the point of intermittency indicating the presence of strong fluctuations. Fig. \ref{MaxDimQ_Plot_2} (a) plots the behavior of the average $\mathrm{max (dim)}\, Q^{i}$ (averaged over $20$ initial conditions), together with the maximum and minimum value seen at each $\mu$. Fig. \ref{MaxDimQ_Plot_2}(b) plots the average value of $\mathrm{max (dim)}\, Q^{i}$ together with its standard deviation.

So far, we discussed the results of the six topological characterizers that was used to analyse the time series networks for various dynamical regimes of the logistic map. We shall now subject these dynamical regimes to a more standard analysis using the conventional complex network characterizers in the next section.

%--------------------------------------------------------------------------------------------------------------------------------------------
% MAXDIMQ PLOT 1
%--------------------------------------------------------------------------------------------------------------------------------------------
%--------------------------------------------------------------------------------------------------------------------------------------------

%--------------------------------------------------------------------------------------------------------------------------------------------
% MAXDIMQ PLOT 2
%--------------------------------------------------------------------------------------------------------------------------------------------

%--------------------------------------------------------------------------------------------------------------------------------------------

%--------------------------------------------------------------------------------------------------------------------------------------------
%--------------------------------------------------------------------------------------------------------------------------------------------
\section{Using Conventional Network Characterizers\label{ConventionalCharacs}}

In this section, we employ three of the widely used conventional complex network characterizers, viz. the average clustering coefficient $c$, the characteristic path length $l$ and the degree distribution \cite{Strogatz98, BA99}, to carry out the standard analysis of the network properties of the TS networks obtained  from the logistic map time series for different dynamical regimes. 
We use the same time series data as used for the simplicial characterization, viz. the logistic map time series for $10,000$ time steps after discarding $5,000$ initial transients, and the resultant TS network of $10,000$ nodes. The three standard network characterizers, viz. the clustering co-efficient, the characteristic path length, and the degree distribution are defined in the next section. We first examine these characterizers separately, and then examine them together.

%--------------------------------------------------------------------------------------------------------------------------------------------
\subsection{Clustering Coefficient}

First, the average clustering coefficient $c$ of a network measures  the extent to which  a network is interconnected, i.e. to what extent are the neighbors of a given node, neighbors of each other.
If a node $i$ is connected to $k$ other nodes, then the clustering coefficient $c_i$ of the node $i$ is defined as the ratio of the actual number of connections that exist between the $k$ nodes to the maximum number of interconnections that can exist between the $k$ nodes. If the actual number of interconnections that exist between $k$ nodes that are linked to the node $i$ is given by $E_i$, then the maximum number of interconnections possible between the $k$ nodes is simply $C^k_2$ which is $k (k - 1) /2$. 
Thus, the clustering coefficient of node $i$ is defined as $c_i$ = $2E_i / k ( k - 1)$.

We first examine the clustering coefficient $c$ for TS networks of different dynamic regimes. 
In the Table \ref{Stand_Characs_Table}, the second column lists  the clustering coefficient spanning over the entire range of values of $\mu$. We see that these values of $c$ that correspond to different dynamical regimes -  periodic, intermittent and chaotic dynamics - all fall in a narrow range, $0.6913 < c < 0.7858$. Each period has a network characteristic of its own period. The clustering coefficients for different periods, however, differ by very small values. For high periods, the tolerance to which the visibility condition is evaluated, also plays a role. The values for intermittent, and chaotic regimes do not fall within distinct ranges.
Thus, the clustering coefficient $c$ is neither able distinguish between distinct dynamic  regimes nor is able to club together similar regimes. 

%--------------------------------------------------------------------------------------------------------------------------------------------
\subsection{Characteristic Path length}

The second characterizer that we look at, is the average path length $l$ of the 
TS-network (see Table \ref{Stand_Characs_Table}), i.e.\,the average distance between arbitrarily chosen points. Here, the average path length $l$ takes  larger
values  in periodic regimes (i.e.\,at $\mu= 3.5, 3.836$), and at the edge of chaos (Feigenbaum point, $\mu=3.56995$), 
than that at the $\mu$ values corresponding to intermittency ($\mu=3. 8284, 3.857$).  Small values of the average path lengths  are seen at the chaotic values $\mu=3.87$, and $\mu=3.89$  and at the fully developed chaos value $\mu=4.0$.
The reason for the existence of large path lengths in the periodic regime is clear. The periodic networks have simplices which connect among near neighbors, and many short steps are necessary to connect points which are far apart on the time series. In terms of the simplicial analysis above, the network structures corresponding to 
periodic orbits have a significantly large number of regular simplices that are connected at lower topological levels. For instance, for $\mu=3.5$ and $\mu=3.836$ the first structure factor $Q(q)$ has a large component at the $q=1$ level, $Q = 501 $ and $Q=667$, respectively. This means that the resultant network is sparsely interconnected, and its average path length is large. 

On the other hand, for the intermittent and chaotic cases, the corresponding TS networks
contain links which connect points which are widely separated on the network. These long range links imply that widely separated nodes can be reached in far fewer steps, leading to short average path lengths. 
In the simplicial language, the chaotic TS networks 
have a large number of regular simplices, that are rather sparsely 
connected at lower topological level ($q=1$, $q=2$), but are 
better connected at a higher topological level ($q=2$ and higher).
The resulting network therefore is an overall  better interconnected network giving rise to a significantly low average path length.

It is interesting to speculate whether there are any regimes which have a small world connectivity, e.g. at the Feigenbaum point or the edge of chaos. However, the present data does not permit any definite conclusion. This question will be examined elsewhere.

%--------------------------------------------------------------------------------------------------------------------------------------------
% STANDARD CHARACS TABLES
%--------------------------------------------------------------------------------------------------------------------------------------------
\begin{table*}[!t]
\setlength{\tabcolsep}{12pt}
\caption{The conventional characterizers - the clustering coefficient and the average path length - for parameters of the logistic map corresponding to various dynamical regimes. In generating the TS-network, we discarded 5\,000 initial transients and used 2\,000 time steps in the time series, thereby the TS-network has nodes N = 2\,000. Calculations of these characterizers was made using NetworkX, a Python language package for network analysis \cite{networkx}
%The conventional characterizers - the clustering coefficient and the characterisitic path length - for parameters
%of the logistic map corresponding to various dynamical regimes. In generating the TS network, we discarded $5,000$
%initial transients and used $10,000$ time steps in the time series, thereby the TS network has nodes $N=10,000$. 
\label{Stand_Characs_Table}}
\centering
\begin{tabular}{ *{5}{c} }
\hline
\hline
$\mu$	&	C	&	L	&	$\lambda$	&	Nature of Orbit	\\
\hline
3.5	&	0.6913	&	167.9995	&	-0.8725	&	Period 4	\\
3.836	&	0.7998	&	222.9999	&	-0.2306	&	Period 3	\\

3.56995	&	0.6932	&	44.2464	&	0.0050	&	Feigenbaum Point	\\

3.8284	&	0.7811	&	193.5178	&	0.1109	&	Intermittency before P3	\\
3.857	&	0.7760	&	46.4710	&	0.2766	&	Intermittency	\\

3.87	&	0.7317	&	34.0742	&	0.4265	&	Chaos 1	\\
3.89	&	0.7320	&	29.1327	&	0.4982	&	Chaos 2	\\
4	&	0.7858	&	27.4089	&	0.6931	&	Full chaos	\\
\hline
\end{tabular}

\end{table*}
%--------------------------------------------------------------------------------------------------------------------------------------------

%--------------------------------------------------------------------------------------------------------------------------------------------
\subsection{Degree distributions}

\begin{figure*}[!t]
	\includegraphics[width=\textwidth]{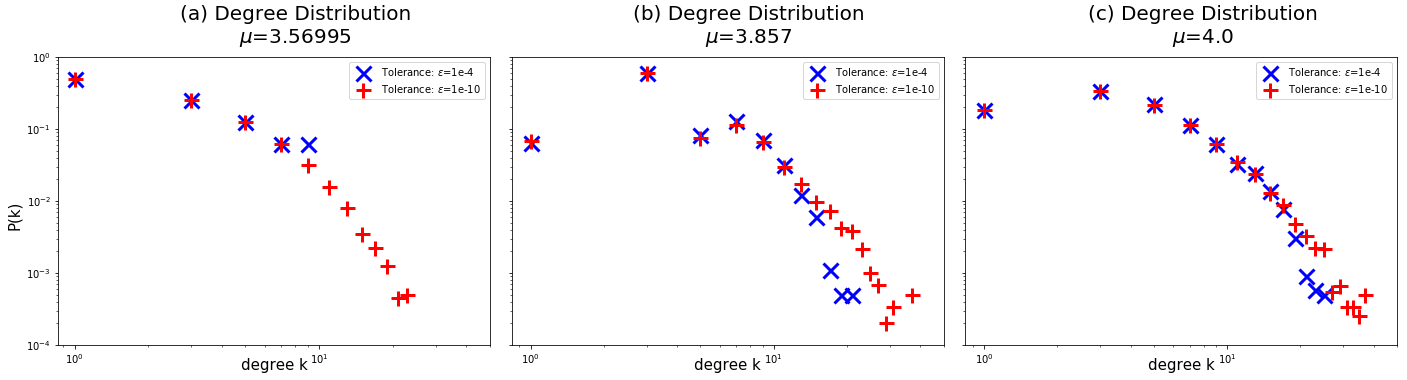}
	\captionsetup{justification=raggedright, singlelinecheck=false}
	\caption{Degree distributions for logistic map TS-networks at (a) $\mu = 3.56995$ (Feigenbaum point), (b) $\mu = 3.857$ (intermittency), and (c) $\mu = 4.0$ (full chaos). Length of TS = $2\,000$.\label{DegDistsFig}}
\end{figure*}

The degree distributions of the TS networks obtained at the $\mu$ values of interest are plotted in the Figure \ref{DegDistsFig}.  It is clear that the periodic networks have many nodes whose interconnections follow identical patterns, and hence there are only a finite number of degrees. This can be seen for $\mu=3.450$ (period 4), and $\mu=3.82843$, the onset of period 3.  
The degree distribution starts showing a bigger variation, over one decade, at the onset of chaos ($\mu=3.56995$). This expands further at $\mu=3.85700$, at the onset of crisis induced intermittency, and  even more so at the chaotic value $\mu= 3.88000$ and $\mu=4.0$, i.e. at fully developed chaos. The log-log plots of these distributions show short regimes where a power-law can be fitted. However, even the $10\,000$ node networks do not really show scale free behavior.

%--------------------------------------------------------------------------------------------------------------------------------------------
% DEG DIST PLOT
%--------------------------------------------------------------------------------------------------------------------------------------------
%--------------------------------------------------------------------------------------------------------------------------------------------

%--------------------------------------------------------------------------------------------------------------------------------------------

We note that our description is capable of identifying network motifs, and is in fact more general than what is provided by network motifs, since it can identify the ways in which the network motifs are put together, the regularity with which they occur, and also motifs at different levels of topological complexity. We hope to discuss this in more detail in future work. 

%--------------------------------------------------------------------------------------------------------------------------------------------
%--------------------------------------------------------------------------------------------------------------------------------------------
\section{Comments on Computational Issues \label{comp_issues}}

We note that the details of the TS-networks show some sensitivity to the accuracy of computation. For example, the visibility condition is evaluated to some tolerance $\epsilon$. This condition makes a difference to the details of the network in some cases, particularly in the case of the Feigenbaum attractor, and leads to small changes in the values of the topological characterisers, especially the exact value of max(dim $Q^{i}$). Similarly, different initial conditions also lead to slightly different values of the topological characterisers, and therefore averages over initial conditions lead naturally to non-integer values of the topological characterisers. We note however, that the changes are small, and the qualitative behaviour of the topological quantities as functions of the level $q$ maintain the behaviour that is shown in the graphs, with each kind of behaviour being characteristic of the given dynamical regime. We also note that these differences reduce with increase in the chaoticity of the system. 

%--------------------------------------------------------------------------------------------------------------------------------------------
%--------------------------------------------------------------------------------------------------------------------------------------------
\section{Conclusion \label{conclusion}}

To summarise, we examine the TS networks obtained from the time series of the logistic map using algebraic topology methods. Our characterisers are clearly able to distinguish between chaotic and periodic regimes. Both regimes contain graphs whose simplicial structure contain nodes, links and triangular faces, and also contain fully connected clique complexes. The periodic regimes are characterised by regular graphs and fewer simplicial structures of dimensions one, two, and three. In contrast, the simplicial structures in the chaotic regimes contain many more connections at higher levels upto nine. The entropies at the highest topology level are higher for the periodic cases, and are significantly lower for the chaotic cases, indicating that the chaotic regimes have higher complexity.

While the dynamical stability of dynamical systems is well understood, and quantified nicely by the Lyapunov exponent, the short term correlations of evolving systems, especially in the chaotic regime, have not been quantified to any great extent. The TS networks constructed by the visibility method encode these correlations in terms of the connectivity of the network graphs. The simplicial characterizers uncover the hidden geometry of these graphs, level by simplicial level, by providing a precise quantification of the manner in which these graphs are connected, pointwise, linkwise, trianglewise, and higher. This is very clear from our tables, as well as from the graphs. This is analogous to the manner in which the multifractal structure 
analyses the scaling behavior of a multiscale set. To the best of our knowledge, there is no analysis of the short term correlations in an evolving system which does this, including entropic analysis. The algebraic characterizers, and their identification of the hierarchy of geometrical structures can also contribute to the identification of dominant network motifs, in general complex networks.

We note that the local quantities pick up the differences in the two cases most sharply, especially the maximum dimension which counts the number of simplices in which the most highly connected node participates. The utility of the algebraic topological quantifiers is thus demonstrated in a simple context where the dynamical behaviour is well understood. Hence, they look like promising candidates for revealing the hidden geometry of networks which represent time series with nontrivial correlations between dynamical states. We expect them to be particularly useful in situations which exhibit phase transitions or other radical changes, such as crisis, intermittency and unstable dimension variability. We hope our study will motivate future work in these directions.

We also note that our local topological quantity, viz. the maximum dimension which counts the number of simplices in which the most highly connected node participates is highly sensitive to the dynamic nature of time series, quantifies the strong increase in the connectivity properties of the network seen at the edge of chaos, and in the chaotic regime, very accurately. The utility of the algebraic topological quantifiers is thus demonstrated in a simple context where the dynamical behaviour is well understood. 

A comment about the usual network characterizers is also necessary here. The clustering coefficient is much the same in all the dynamic regimes. This indicates that the clique formation in the distinct dynamic regimes is not significantly different. However, the short  path lengths on the network in the chaotic regime  encode the fact that the connections formed here are long range connections on the network,  as opposed to the long path lengths (and short range connections) in the periodic regime. It is interesting to note that the edge of chaos (the Feigenbaum point) and chaotic regime at the end of the period $3$ window shows path length values which are clearly separated from both these regimes. This small world like behavior is in line with other observations which indicate distinctly different behavior at the edge of chaos.  This point deserves further investigation, and needs to be supplemented by further investigation of the simplicial structure at the edge of chaos.

Thus, the algebraic topological characterizers of time series networks appear to be  promising candidates for revealing the hidden geometry of networks which represent time series with nontrivial correlations between dynamical states. We expect them to be particularly useful in situations which exhibit phase transitions or other significant  changes in the dynamics, such as jamming behavior  and unstable dimension variability. 
We note that our description is capable of identifying network motifs, and is in fact more general than what is provided by network motifs, since it can identify the ways in which the network motifs are put together, the regularity with which they occur, and also motifs at different levels of topological complexity. We hope our study will motivate future work in these directions.

%--------------------------------------------------------------------------------------------------------------------------------------------
%--------------------------------------------------------------------------------------------------------------------------------------------
\begin{acknowledgements}
NG and NNT thank the CSIR scheme (03(1294)/13/EMR-II) for partial support.
\end{acknowledgements}

%--------------------------------------------------------------------------------------------------------------------------------------------
%--------------------------------------------------------------------------------------------------------------------------------------------
\appendix*

%--------------------------------------------------------------------------------------------------------------------------------------------
\section{Calculation of six characterizers using a simple example }

\begin{figure}[b]
\begin{center}
\setlength{\unitlength}{0.012500in}%
	\includegraphics[scale=0.5]{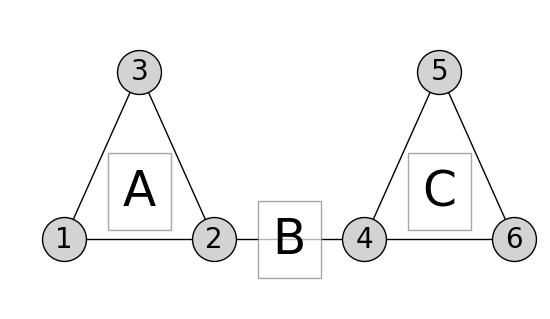}
\end{center}
\captionsetup{justification=raggedright, singlelinecheck=false}
\caption{An illustration to demonstrate connectivity between three simplices in a simplicial complex. Two simplices $A$ and $C$ are of dimension $q=2$, and the third simplex $B$ is of dimension $q=1$. Simplices $A$ and $C$ are $0$-connected to simplex $B$, which means that each of them has a single vertex in common with $B$. We use this example to illustrate the calculation of all the six characterizers in the appendix.
\label{ExampleSimpStruc}}
\end{figure}

Let us take the simplicial complex in Fig. \ref{ExampleSimpStruc} where two triangular simplices that 
are connected by a link. The simplicial complex here has three simplices, the third one being the simplex made of 
two nodes with a single link between them.
We shall now use this simplicial complex to illustrate how the six characterizers are calculated.\\

The three simplices of the simplicial complex are denoted by
$A = \lbrace 1, 2, 3 \rbrace$ and $B = \lbrace 2, 4 \rbrace$, and  $C = \lbrace 4, 5, 6 \rbrace$, with the vertices labelled as shown in the Fig. \ref{ExampleSimpStruc}. The incidence matrix $\mathbf{\Lambda}$  of the simplicial complex
can be written as a matrix with the simplex index as rows, and the node index as columns. If a node
is contained in a simplex then the corresponding element in the matrix is $1$, or else it is $0$. We therefore, have 

$$ \mathbf{\Lambda} = \begin{pmatrix}
                 		1 & 1 & 1 & 0 & 0 & 0\\
                	    0 & 1 & 0 & 1 & 0 & 0\\
                	    0 & 0 & 0 & 1 & 1 & 1\\
                	    \end{pmatrix}$$

Both the simplices $A$ and $C$ are of dimension $2$, as they have $3$ nodes each,
and the simplex $B$ has a dimension $1$, because it has $2$ nodes.
Note that if a simplex has $q+1$ nodes, then it is of dimension $q$. 

Let us first calculate the first structure vector $\mathbf{Q}$ for the simplicial complex. 

%--------------------------------------------------------------------------------------------------------------------------------------------
%--------------------------------------------------------------------------------------------------------------------------------------------

%--------------------------------------------------------------------------------------------------------------------------------------------
\subsection{First Structure Vector $\mathbf{Q}$}

Let us consider the first structure vector $\mathbf{Q}$ for the simplicial complex. Since the simplices $A$ and $B$ are 
with three nodes (their dimension being $2$), the structure vector $\mathbf{Q}$ will
have three levels : $q=0, q=1$ and $q=2$. If two simplices are to be $q$-connected,
they should have at least $q+1$ nodes in common, and also that if two simplices are 
$q$-connected, then they are also connected at all lower topological levels.

In our example, at $q=0$ level, for topological connectivity between
any two simplices we need at least $one$ node in common. By this, we
see both the pairs $A$ and $B$ as well as $B$ and $C$ are connected, as
they have one common node each, node-$2$ and node-$4$, respectively. 
In fact, the whole simplicial complex made up of three simplices $A,B,$ and $C$ is 
now identified as one entity at the $q=0$ topological level. 

At the next level $q=1$, for connectivity between two simplices we need at least
two nodes to be in common. As we have none, all the three simplices are disconnected
from each other at this level. The total number of entities is now three. 

At $q=2$ level, none of the simplices are connected for it requires
a minimum of three nodes to be in common. And in addition, 
at $q=2$ level, only the simplices $A$ and $B$ with dimension $3$
exist. Therefore, we see only two entities at this level. 
In all, the first structure vector is $\mathbf{Q} = (1,3,2)$.

%--------------------------------------------------------------------------------------------------------------------------------------------

\subsection{Second Structure Vector $\mathbf{N_s}$ }

The second structure vector $\mathbf{N_s}$ is defined as follows. The $q$th component
of $\mathbf{N_s}$ is the number of simplices present in the simplicial 
complex at level $q$ and higher. At the level $q=0$, we can see from 
Fig. \ref{ExampleSimpStruc} that the total number of simplices here is three, $n_{0} = 3$.
Next at level $q=1$, the number of simplices is again three, $n_{1} = 3$, and at level $q=2$, the number is two, $n_{2} = 2$.
Therefore, the second structure vector is $\mathbf{N_s} = (3,3,2)$.

%--------------------------------------------------------------------------------------------------------------------------------------------
\subsection{Third Structure Vector $\mathbf{\widehat{Q}}$ }

The $q$th component of the third structure vector
$\mathbf{\widehat{Q}}$, is given by  $1- Q_q/n_q$, where $Q_q$ and $n_q$ are
the $q$th components of the first and second structure
vectors, respectively. So, we can get $\mathbf{\widehat{Q}}= (2/3,0,0)$.

%--------------------------------------------------------------------------------------------------------------------------------------------
\subsection{$\mathbf{\tilde{f}}$ vector}

Here, the $q$th component is the number of simplices at level-$q$. 
At level $q=0$, there are no simplices that have one isolated
node, thereby ${\tilde{f}}_0 = 0$. At the next level $q=1$, 
the only simplex to have two nodes is the simplex $B$, thus ${\tilde{f}}_1 = 2$.
For level $q=2$, two simplices $A$ and $C$ have three nodes each, and we have
${\tilde{f}}_2 = 2$. We can now write the vector as $\mathbf{\tilde{f}} = (0,1,2)$.

%--------------------------------------------------------------------------------------------------------------------------------------------
\subsection{Maximum topological dimension: $\mathrm{max (dim)}\, Q^{i}$}

To find the maximum topological dimension of simplicial complex,  we the
need to find the topological dimension $Q^i$ of all the nodes. The
topological dimension of a node $i$ is the number of simplices of dimension-$q$ (same as level-$q$)
the node participates in. For the node-$1$, it takes part only in the simplex
$A$ which is at level-$2$. Therefore, the only 
non-zero component of the vector $\mathbf{Q^1}$ is for the
level-$2$, $\mathbf{Q}^1 = (0,0,1)$.
Similarly, the node-$2$ participates in two simplices $A$, of dimension $2$,
and $B$, of dimension-$1$. Therefore, the vector $\mathbf{Q^1}$ will
have non-zero contributions for levels-$1$ and $2$, giving $\mathbf{Q}^2 = (0,1,1)$.

Working out in a similar fashion, we will have the vectors for the four other
nodes in the simplicial complex as follows. $\mathbf{Q}^3 = (0,0,1), \mathbf{Q}^4 =(0,1,1),
\mathbf{Q}^5 = (0,0,1),\mathbf{Q}^6 = (0,0,1)$.

\subsection{Entropy $\mathbf{S}$ }

The entropy of a topological level-$q$ is defined by 
\begin{equation*}
\mathbf{S} (q) = \frac{-\sum_i\, p_q^i \log p_q ^i}{\log N_q},
\end{equation*}

where, $p_q^i$ is the occupation probability of a node at 
the level-$q$, given by $Q_q^i/\sum_i\ Q_q^i$. The quantity
$N_q = \sum_i\ (1 - \delta_{Q^i_q , 0}) $, is the number of nodes
which have non-zero entry at the level-$q$ in the simplicial complex.
The delta function in $N_q$ will take the value
of unity if the subscript $Q^i_q = 0$, else it will be zero. 

Now, the occupation probability $p_i$ is obtained from the topological dimensions $Q_q^i$ 
(calculated above). At level $q=0$, $Q_q^i$ is zero for all the nodes, thereby $p_i = 0$
and entropy $S_Q(0) = 0$.
At level $q=1$, only the nodes $2$ and $4$ contribute to the topological dimension. That is,
$Q_1^2 = 1$ and    $Q_1^4 = 1$, all else are zero. The corresponding occupation probabilities 
are $p_1^2 = 1/2$ and $p_1^4 = 1/2$. The entropy at level $q=1$ is $S(1) = 0.8908$.

Finally at level $q=2$, all the nodes contribute to the topological dimension, $Q_2^i = 1$, 
that gives a values of $p_2^i = 1/6$. Then, the entropy at the level $q=2$ is
$S(2) = 1$.

The quantities defined here can now be computed for the actual TS-networks in a similar way.

%--------------------------------------------------------------------------------------------------------------------------------------------
%--------------------------------------------------------------------------------------------------------------------------------------------
%\begin{thebibliography}{}

\bibliographystyle{apalike}

\begin{thebibliography}{}

\end{thebibliography}


\begin{thebibliography}{}
%1
\bibitem[Kantz and Schreiber, 2004]{Kantz}
H. Kantz and T. Schreiber,
{\em Nonlinear Time Series Analysis}
(Cambridge University Press, New York, 2004).


%2
\bibitem[Abarbanel, 1996]{TSA_refs}
H. D. I. Abarbanel,
{\em Analysis of Observed Chaotic Data}
(Springer-Verlag, New York, 1996);
B. Schelter et al.,
{\em Handbook of Time Series Analysis}
(Wiley-VCH, 2006).

%3
\bibitem[Gao, Small, and Kurths, 2016]{GaoEPL2016}
Z. Gao, M. Small, and J. Kurths,
EPL \textbf{116}, 50001
(2016).


%4
\bibitem[Lacasa et al, 2008]{vis_algo}
L. Lacasa, B. Luque, F. Ballesteros, J. Luque, and J. C. Nu\~{n}o,
Proc. Nat. Acad. Sci. USA
\textbf{105}, 4972
(2008).

%5
\bibitem[Campanharo, et. al, 2011]{Campanharo}
A. S. L. O. Campanharo, M. I. Sirer, R. D. Malmgren, F. M. Ramos, and L. A. N. Amaral,
PloS One
\textbf{8}, e23378
(2011).

%6
\bibitem[Marwan, et. al, 2009]{Marwan}
N. Marwan, J. F. Donges, Y. Zou, R. V. Donner, and J. Kurths,
Phys. Lett. A
\textbf{373}, 4246
(2009).

%7
\bibitem[Strogatz, 1998]{Strogatz98}
D. J. Watts and S. H. Strogatz,
Nature
\textbf{393}, 440
(1998).

%8
\bibitem[Albert, Barabasi, 2002]{AlbertBarabasi2002}
R. Albert and A-L. Barab{\'a}si,
Rev. Mod. Phys
\textbf{74}, 47
(2002).

%9
\bibitem[Kramer and Laubenbacher, 1998]{Kramer}
X. H. Kramer and R. C. Laubenbacher,
in
\textit{Proceedings of Symposia in Applied Mathematics},
San Diego, 1998,
edited by D. A. Cox and B. Sturmfels,
(San Diego, California, 1997)
\textbf{53}, p. 93.

%10
\bibitem[Atkin, 1972]{Atkin72}
R. H. Atkin,
Int. J. Man. Mach. Stud.
\textbf{4}, 139
(1972).


%11
\bibitem[Duckstein and  Nobe, 1997]{Duckstein}
L. Duckstein and S. A. Nobe,
Eur. J. Oper. Res.
\textbf{103}, 411
(1997).

%?
\bibitem[Carlsson, 2009]{Carlsson2009}
G. Carlsson,
Bull. Amer. Math. Soc.
\textbf{46}, 255
(2009).

%?
\bibitem{CompTopol}
H. Edelsbrunner and J. Harer,
\textit{Computational Topology: An Introduction}
(AMS, 2010).


%12
\bibitem{Andjelkovic1}
M. Andjelkovi{\'c}, N. Gupte, and B. Tadi{\'c},
Phys. Rev. E
\textbf{91}, 052817
(2015).

%13
\bibitem[Mittal and Gupta, 2017]{MittalGupta2017}
K. Mittal and S. Gupta,
Chaos \textbf{27}, 051102
(2017).

%14
\bibitem[S. Maleti{\'c} et al., 2016]{MaleticRajkovic2016}
S. Maleti{\'c}, Y. Zhao, and Milan Rajkovi{\'c},
Chaos \textbf{26}, 053105
(2016).

%15
\bibitem[Giusti, Ghrist, and Bassett, 2016]{GiustiGhristBassett2016}
C. Giusti, R. Ghrist, and D. S. Bassett,
J Comput Neurosci \textbf{41}, 1
(2016).

%16
\bibitem[Luque, et. al, 2011]{Luque}
B. Luque, L. Lacasa, F. J. Ballesteros, and A. Robledo,
PLoS One
\textbf{6}, e22411
(2011).

%17
\bibitem[Zhang and Small, 2006]{Zhang2006}
J. Zhang and M. Small, 
Phys. Rev. Lett.
\textbf{96}, 238701
(2006).

%18
\bibitem[Yang and Yang, 2008]{Yang2008}
Y. Yang and H. Yang,
Physica A
\textbf{387}, 1381
(2008).

%19
\bibitem[Xu et al., 2008]{SmallSuperfamilyMotifs}
X. Xu, J. Zhang, and M. Small,
Proc. Nat. Acad. Sci. U.S.A. \textbf{105}, 19601
(2016).

%20
\bibitem[Lacasa, 2014]{LacasaNonlin}
L. Lacasa,
Nonlinearity \textbf{27}, 2063
(2014).

%21
\bibitem[Gutin, Mansour, Severini, 2011]{Gutin2011}
G. Gutin, T. Mansour, and S. Severini,
Physica A \textbf{390}, 2421
(2011).

%22
\bibitem[Yang, 2009]{Yang}
Y. Yang, J. Wang, H. Yang, and J. Mang,
Physica A
\textbf{388}, 4431
(2009).

%23
\bibitem[Elsner, 2009]{Elsner}
J. B. Elsner, T. H. Jagger, and E. A. Fogarty,
Geophys. Res. Lett.
\textbf{36}, L16702
(2009).

%24
\bibitem[Maleti{\'c} and  Rajkovi{\'c}, 2012]{Maletic}
S. Maleti{\'c} and M. Rajkovi{\'c},
Eur. Phys. J. Spec. Top.
\textbf{212}, 77
(2012).

%25
\bibitem[Bron and Kerbosch, 1973]{Bron}
C. Bron and J. Kerbosch,
Commun. A.C.M.
\textbf{16}, 575
(1973).

%26
\bibitem[Andjelkovi{\'c}, et. al, 2014]{Andjelkovic}
M. Andjelkovi{\'c}, B. Tadi{\'c}, S. Maleti{\'c}, and M. Rajkovi{\'c},
Physica A
\textbf{436}, 582
(2015).

%27
\bibitem{Jonsson}
J. Jonsson,
\textit{Simplicial Complexes Of Graphs},
(Springer-Verlag Berlin Heidelberg, 2008).

%28
\bibitem[Barabasi, 1999]{BA99}
A. L. Barab{\'a}si and R. Albert,
Science
\textbf{286}, 509
(1999).

%29
\bibitem[Hagberg et al., 2008]{networkx}
A. A. Hagberg, D. A. Schult, and P. J. Swart,
in
\textit{Proceedings of the 7th Python in Science Conference (SciPy2008), Pasadena 2008},
edited by G. Varoquaux, T. Vaught, and J. Millman, p. 11.



\end{thebibliography}
%--------------------------------------------------------------------------------------------------------------------------------------------
\end{document}